\documentclass[aps,prd,reprint,a4paper,showpacs,nofootinbib,superscriptaddress]{revtex4-2}

\usepackage{amsmath}
\usepackage{amssymb}
\usepackage{graphicx}
\usepackage{float}
\usepackage{multirow}
\usepackage{placeins}
\usepackage{bm}
\usepackage{dcolumn}
\usepackage[utf8]{inputenc}
\usepackage[T1]{fontenc}
\usepackage{booktabs}
\usepackage{txfonts}
\usepackage{mathrsfs}
\usepackage{xcolor}
\usepackage{orcidlink}
\usepackage{hyperref}
\hypersetup{colorlinks=true, linkcolor=red, citecolor=blue}
\usepackage{geometry}
\begin{document}

\title{Chaos Bound, Integrability Breaking and Dynamical Indicators for Charged Particles in Rotating 4D Einstein--Gauss--Bonnet Black Hole Backgrounds}

\author{Rana Muhammad Zulqarnain }
\email{ranazulqarnain7777@gmail.com}
\affiliation{School of Business, Xian International University, Xian, 710077, Shaanxi, China.}

\author{Anum Fayyaz}
\email{anamfayyaz1122@gmail.com}
\affiliation{Department of Mathematics, The Islamia University of Bahawalpur, Bahawalpur-63100, Pakistan}

\author{Shubham Kala\orcidlink{0000-0003-2379-0204}}
\email{shubhamkala871@gmail.com}
\affiliation{Centre for Space Research, North-West University, Potchefstroom 2520, South Africa}

\author{G. Abbas}
\email{ghulamabbas@iub.edu.pk}
\affiliation{Department of Mathematics, The Islamia University of Bahawalpur, Bahawalpur-63100, Pakistan}
\affiliation{Jadara University Research Center, Jadara University, Jordan}

\author{S. K. Maurya }
\email{sunil@unizwa.edu.om }
\affiliation{Department of Mathematical and Physical Sciences, College of Arts and Sciences, University of Nizwa, 616, Nizwa, Sultanate of Oman}

\author{G. Mustafa\orcidlink{0000-0003-1409-2009}}
\email{gmustafa3828@gmail.com (Corresponding Author)}
\affiliation{Department of Physics, Zhejiang Normal University, Jinhua 321004, China}

\begin{abstract}
We study a charged test particle around the rotating charged black hole of four-dimensional Einstein--Gauss--Bonnet (EGB) gravity and ask how the coupling $\beta$, the black hole charge $Q$, spin $a$ and the particle angular momentum $L$ control its instability and chaos. The Hamilton--Jacobi equation separates for any metric function $\Delta(r)$, so the motion admits a Carter constant and is integrable; the Lyapunov exponent $\lambda$ of equilibrium orbits thus measures local instability, not chaos. We compute $\lambda$ from the Jacobian of the radial phase space with the exact energy Hamiltonian, which retains the frame-dragging term $\omega L$, and compare it with the surface gravity $\kappa$ through $\lambda\le\kappa$, the classical-probe form of the Maldacena--Shenker--Stanford bound. For non-extremal black holes, co-rotating orbits satisfy the bound in all cases examined, whereas particles with $L\le0$, mainly counter-rotating ones, violate it once $a$, $Q$ or $\beta$ is large enough. For extremal black holes ($\kappa=0$) the bound is not applicable, and $\lambda^2$ serves only as an instability diagnostic: counter-rotating particles have unstable equilibria, co-rotating ones none. To study genuine chaos, we break integrability with a harmonic trap that confines the particle near the horizon. Poincar\'e sections, the largest Lyapunov exponent and the smaller alignment index show that chaos grows as the energy approaches its horizon value, with exponents of $0.13$--$0.22\,\kappa$, although the local exponent at the near-horizon equilibrium of the trap equals $\kappa$ to within $10^{-5}$. A control without the black hole shows that the trap itself supports part of the chaos, which the black hole redistributes towards the horizon. We state the limits of this test-particle, effective-background setting.\\

\end{abstract}

\maketitle

\section{Introduction}

Even while General Relativity (GR) has been incredibly successful and elegant, it is generally considered to be an inadequate fundamental description of gravity. GR is frequently seen in the current framework as a low-energy effective theory, indicating the need for a more complete theory that can explain gravitational phenomena outside of the classical regime. EGB gravity has become a particularly attractive option for modified theories of gravity among the different extensions of general relativity. Interestingly, the EGB framework emerges naturally as an efficient description at string theory's low-energy limit, offering a strong theoretical justification for investigating its gravitational implications \cite{Zwiebach1985,Garraffo2008}. The GB invariant becomes topological in four spacetime dimensions in the traditional formulation of EGB gravity. As a result, it reduces to a total derivative and does not directly contribute to the local gravitational field equations. This characteristic typically keeps the gravitational dynamics in $(D=4)$ from being altered by the GB term. Higher-curvature corrections to four-dimensional gravitational phenomena can now be investigated according to Glavan and Lin's new prescription, which results in a non-trivial four-dimensional formulation of EGB gravity \cite{Glavan2020}.
Glavan and Lin developed a rescaling of the Gauss-Bonnet coupling constant $(\alpha \rightarrow \alpha/(D-4))$, followed by the limiting approach $(D\rightarrow4)$ to get a non-trivial four-dimensional limit. The Gauss-Bonnet contribution can continue to be dynamically important in four dimensions because to this prescription. Surprisingly, the resulting theory offers a new framework that avoids the Ostrogradsky instability linked to higher-derivative gravitational theories while getting around the traditional limitations imposed by Lovelock's theorem. In \cite{Glavan2020}, a static and spherically symmetric (SSS) black hole solution was later obtained in this framework.

An interesting way to investigate the complex gravitational structure of black hole (BH) spacetimes is through the motion of test particles. Understanding nonlinear dynamics and the related dynamical instabilities that arise in such systems has been the focus of significant study over the past few decades. To detect and describe chaotic behavior in gravitational environments, a number of complementary analytical and computational techniques, such as basin-boundary analysis, Lyapunov-exponent methods, and Poincaré sections, have been widely used  \cite{Cornish:1996ri,Bombelli:1991eg,Zayas:2010fs}. The study of BH chaos was inspired by more developments in the AdS/CFT correspondence and the research of quantum information scrambling \cite{Hashimoto2017Universality,Hashimoto:2018fkb}. In this context Maldacena \textit{et al.} \cite{Maldacena:2015waa} made the proposal that the surface gravity $\kappa$ at the event horizon should be greater than or equal to the classical Lyapunov exponent $\lambda$ of the chaotic motion.

In many branches of physics, especially gravitational systems, where large spacetime curvature leads to rich nonlinear dynamics, this inherent complexity has garnered significant attention. In black hole physics, the paths of test particles in progressively complex spacetime geometries are frequently used to study chaotic behavior. Charged particle motion around black holes in quasi-topological electrodynamics \cite{Lei2021Chaos}, test particle dynamics in deformed electrically and magnetically charged Reissner--Nordström black holes \cite{Bokhari2020TestParticles}, and particle motion in the Rindler-modified spacetime of four-dimensional spherically symmetric black holes \cite{Guo2021Minimal} have all been studied recently. These studies show that the beginning and nature of chaotic motion are significantly influenced by the geometric and electromagnetic aspects of the underlying spacetime.

The chaotic motion of particles and strings near rotating and non-rotating black holes has been thoroughly studied in recent years. Charged particles and the corresponding chaos bound in various black hole spacetimes have received special study. Variations in black hole charge, spin, cosmological constant, scalar coupling, and particle angular momentum have revealed varied chaotic behaviors. These investigations demonstrate that the stability and intensity of chaos are strongly influenced by both spacetime geometry and particle characteristics.
With an emphasis on Lyapunov exponents and chaos-bound violations, recent research has investigated charged-particle dynamics around Einstein–Euler–Heisenberg–AdS, Einstein–Maxwell–Chern–Simons, nonlinear electrodynamics, and Einstein–Maxwell–Dilaton–Axion black holes \cite{ChenGao2022,Yu2023,Gao2026,YuChenGao2022}. These papers show that black hole charge, spin, scalar or Euler–Heisenberg parameters, particle charge, and angular momentum greatly affect chaotic behavior. Chaos-bound violations from near-horizon areas to finite distances have been found by additional research employing Jacobian-matrix approaches \cite{WangChen2023}. Overall, the findings show that spacetime and particle properties have a significant impact on chaotic dynamics.

The chaotic motion of particles and strings near rotating and non-rotating black holes has been studied extensively in recent years, and charged particles and the
corresponding chaos bound have received particular attention. Varying the black hole
charge, spin, cosmological constant and scalar coupling, and the particle angular
momentum, leads to a variety of chaotic behaviours. Recent work on charged-particle
dynamics around Einstein--Euler--Heisenberg--AdS, Einstein--Maxwell--Chern--Simons,
nonlinear-electrodynamics and Einstein--Maxwell--dilaton--axion black holes has
focused on Lyapunov exponents and violations of the chaos
bound~\cite{ChenGao2022,Yu2023,Gao2026,YuChenGao2022}, and has shown that the
black hole charge, spin and scalar or Euler--Heisenberg parameters, together with the particle charge and angular momentum, strongly affect the chaotic behaviour. Using the Jacobian-matrix method, Ref.~\cite{WangChen2023} found violations of the bound both near the horizon and at finite distances from it. Overall, these studies show that both the spacetime and the particle properties have a significant effect on the dynamics.

The appearance of chaos is closely tied to the loss of integrability~\cite{Bombelli:1991eg}. A Hamiltonian
system with as many independent integrals in involution as degrees of freedom is
integrable in the Liouville sense~\cite{audin2012symplectic,Wang:2021gja}. Its bounded orbits lie on invariant tori, its
Poincar\'e sections consist of smooth invariant curves, and its orbits are regular~\cite{Chen:2016tmr,Lei:2020clg,Basu:2012ae,Deich:2022vna}. A positive Lyapunov exponent at an unstable equilibrium or circular orbit of such a
system measures the local instability of that orbit, that is, the rate at which
neighbouring orbits leave it, and does not by itself indicate chaos~\cite{pikovsky1945lyapunov,sandri1996numerical}. Work on relativistic Hamiltonian dynamics has explored the effects of nonlinear interactions on phase-space structure~\cite{el2025chaotic,el2026extended,el2026stability,El-Nabulsi:2026pls}, alongside the use of differential Galois theory to examine integrability in curved spacetimes~\cite{NLD2023curved,Szuminski:2025nld}. Chaos requires a coupling between gravitational, electromagnetic or other background fields, or an
external force, that makes the equations of motion non-separable and removes one of
the integrals. Geodesic and charged-particle motion in the Kerr--Newman family is
integrable because of the Carter constant~\cite{Carter:1968rr}, so chaos in these
backgrounds requires such a perturbation. A well-studied example is a particle
confined near the horizon by a harmonic trap. The motion in the trap alone is regular for small oscillations, and the motion in the black hole background alone is
integrable, but their combination is not, and the exponential instability supplied by the horizon makes the trapped motion chaotic. This was shown for a massless and
electrically neutral particle near black hole and Rindler horizons, with the largest
Lyapunov exponent bounded by the surface gravity~\cite{Dalui2019Presence,Dalui2020Induction},
building on the analysis of particles held near the horizon in Ref.~\cite{Hashimoto2017Universality}. In recent years, this method has been increasingly employed to examine the integrability properties of Hamiltonian systems associated with different black hole spacetimes~\cite{Chen2019BornInfeld,Kala:2026zuf,Kala:20267}.
Chaos can therefore originate either in the complexity of the spacetime and field
configuration or in an external perturbation that modifies the particle dynamics.

The dynamics of a string approaching the event horizon of a charged Kiselev black hole was studied in Ref.~\cite{Li2025StringChaos}, which found strongly chaotic motion with many violations of the MSS bound and showed that the angular momentum and energy of the string affect both the onset and the strength of chaos. In a related work, the chaotic motion of a charged particle in the Kerr--Newman--Taub--NUT spacetime was studied with the Lyapunov exponent computed from the Jacobian matrix, which proved to be a useful measure of orbital stability~\cite{chen2023chaos}. That work found that the particle angular momentum is the parameter that most strongly controls the size of the violation, that in the extremal case the bound was violated throughout the parameter range considered, and that in the non-extremal case the extent of violation depends on the black hole parameters. It also showed that the mass, charge, spin and NUT parameter of the black hole strongly affect the motion, the NUT charge doing so through the gravitomagnetic effects it introduces into the geometry and the horizon structure.

Despite extensive recent work on EGB black holes, the chaotic motion of particles in
rotating EGB spacetimes has received little attention. Motivated by this gap, the
present work studies the dynamics and chaos of charged particles around rotating
charged EGB black holes. The horizon structure, ergosphere, innermost stable circular orbits (ISCOs), effective potential, particle collisions and thermodynamic properties of a charged rotating black hole in 4D EGB gravity were investigated in
Ref.~\cite{nosirov2024particle}. Using the Gauss--Bonnet theorem of differential
geometry in the weak-field limit, Ref.~\cite{li2022deflection} studied the
gravitational deflection of charged particles by a four-dimensional charged EGB black hole and the dependence of the deflection angle on the GB coupling, the black hole charge, the particle charge-to-mass ratio and the particle velocity. Using the
Newman--Janis algorithm, Ref.~\cite{papnoi2022rotating} constructed a charged rotating black hole in 4D EGB gravity and examined its horizon structure, shadow, effective potential and energy emission rate, showing how the GB coupling and the electric charge affect the shadow and the spacetime properties. Since the Newman--Janis algorithm does not in general produce solutions of modified gravity theories~\cite{Hansen:2013owa}, we treat this metric as an effective
background rather than as an exact solution of a particular four-dimensional theory
(Sec.~\ref{s5}).

Chaos of charged probes has been studied in many rotating
backgrounds~\cite{fayyaz2025chaotic,fayyaz2025massive,fayyaz2026phase,chen2023chaos,Kala:2026qej,Singh:2026rbz,Kala:2026etg}, but the effect of a higher-curvature coupling on the instability and chaos of charged particles in a \emph{rotating charged} geometry has received little attention. The rotating charged EGB black hole of
Refs.~\cite{kumar2020rotating,papnoi2022rotating,nosirov2024particle} is a natural
setting for this question: it has a single additional parameter, the GB coupling
$\beta$, and reduces to the Kerr--Newman black hole as $\beta\rightarrow0$, so the
role of the coupling can be isolated cleanly.

We first show that the Hamilton--Jacobi equation of a charged particle separates in
this background for any metric function $\Delta(r)$, so that the motion admits a
Carter constant and is integrable; Poincar\'e sections of orbits near the ISCO
accordingly consist of smooth invariant curves for all values of $\beta$, $Q$ and $a$ considered. The Lyapunov exponent of an equilibrium orbit therefore measures local instability, and its comparison with $\kappa$ is a statement about this local
exponent rather than about chaos. We obtain the exponent from the Jacobian matrix of
the radial phase space, following Ref.~\cite{chen2023chaos}, but we use the exact
energy Hamiltonian in the frame-dragging frame, which retains the frame-dragging term $\omega L$; omitting this term changes the equilibria and the exponents for $L\neq0$. We claim no novelty for the Jacobian-matrix method, and violations of $\lambda\le\kappa$ by probe particles and strings are already known in many
backgrounds~\cite{ChenGao2022,YuChenGao2022,Yu2023,Gao2026,WangChen2023,chen2023chaos,Li2025StringChaos}.

To study genuine chaos, we break the integrability with a harmonic trap of the form
used in Ref.~\cite{Dalui2019Presence}, which confines the particle near the horizon
and destroys the Carter constant. We characterise the trapped motion with Poincar\'e
sections, the largest Lyapunov exponent (LLE)~\cite{benettin1980lyapunov} and the
smaller alignment index (SALI)~\cite{Skokos:2004mt}. At the energies needed to reach
the horizon the trap is far from its small-oscillation regime and the motion is
non-separable even without the black hole, so we compare the results with a control
calculation in flat spacetime. Finally, we compare the trajectory LLE with $\kappa$
and with the local exponent at the near-horizon equilibrium created by the trap,
which connects the two parts of the paper.

Our contribution is a systematic characterisation of how $\beta$, $Q$, $a$ and $L$
jointly control the local instability and the chaos of charged particles in this
geometry, with co-rotating and counter-rotating orbits treated separately. For
non-extremal black holes, co-rotating orbits satisfy $\lambda\le\kappa$ in all cases
examined, whereas particles with $L\le0$, mainly counter-rotating ones, violate it once $a$, $Q$ or $\beta$ is sufficiently large. For the trapped particle, chaos grows as the energy approaches its value at the horizon, with largest Lyapunov exponents between $0.13\kappa$ and $0.22\kappa$, although the local exponent at the near-horizon equilibrium of the trap equals $\kappa$ to within $10^{-5}$. The flat-spacetim control shows that part of this chaos is supported by the trap itself, so the black hole mainly redistributes it towards the horizon. 

At extremality $\kappa=0$, so $T=0$ and the thermal derivation of the bound does not
apply; the $\kappa$-form of the bound then demands $\lambda^2\le0$, which any unstable orbit fails trivially. We therefore use extremal black holes only as a diagnostic of instability, through the magnitude of $\lambda^2$ and the existence of unstable equilibria, and as the $\kappa\rightarrow0$ limit of the non-extremal analysis, not as an independent violation. In this limit counter-rotating particles have unstable equilibria, whereas co-rotating ones have no equilibrium outside the horizon.

The paper is organised as follows. Section~\ref{s2} introduces the spacetime geometry, electromagnetic potential, horizon structure, and surface gravity. Section~\ref{S3} develops the charged-particle Hamiltonian formulation, establishes the separability of the equations of motion, and derives the corresponding local Lyapunov exponent. Section~\ref{s3} investigates the chaos bound and its possible violation for rotating Einstein--Gauss--Bonnet black holes, presenting results for both non-extremal configurations and extremal black holes. Section~\ref{S4} introduces the harmonic trap and examines the resulting non-integrable dynamics using Poincar\'e sections, the largest Lyapunov exponent (LLE), and the smaller alignment index (SALI). In Section~\ref{s_disc}, the local exponents of the equilibrium orbits are compared with the trajectory LLE of the trapped particle, which connects the two parts of the analysis. Finally, Sec.~\ref{S5} summarises the main findings and conclusions, and states the assumptions and limitations of the present setup.

\section{Rotating  Einstein-Gauss-Bonnet black holes }\label{s2}
\subsection{Review of Rotating  Einstein-Gauss-Bonnet black holes }
The Einstein--Maxwell--Gauss--Bonnet (EMGB) action in a $D$-dimensional spacetime can be written as 
\begin{equation}
S=\frac{1}{16\pi}\int d^{D}x\,\sqrt{-g}
\left[
R+\frac{\beta}{D-4}\mathcal{G}
-F_{\mu\nu}F^{\mu\nu}
\right],
\label{ee1}
\end{equation}
 in accordance with \cite{fernandes2020charged}. In this case, $g$ is the metric tensor determinant, $\beta$ is the Gauss-Bonnet coupling constant, and $F_{\mu\nu}$ is the electromagnetic field-strength tensor. Where $R$, $R_{\mu\nu}$, and $R_{\mu\nu\rho\sigma}$ represent the Ricci scalar, Ricci tensor, and Riemann curvature tensor, respectively, the Gauss--Bonnet invariant is defined as \cite{fernandes2020charged}
 \begin{equation}
\mathcal{G}
=
R^{2}
-4R_{\mu\nu}R^{\mu\nu}
+R_{\mu\nu\rho\sigma}R^{\mu\nu\rho\sigma},
\label{}
\end{equation}

Using the Newman-Janis method and the modified non-complexification prescription suggested by Azreg-Aïnou \cite{azreg2014generating,azreg2014static}, Kumar and Ghosh \cite{kumar2020rotating} built the rotating Einstein--Gauss--Bonnet (EGB) black hole solution. Using the static, spherically symmetric solutions found by Glavan and Lin \cite{Glavan2020}, this method has been successful in producing rotating imperfect-fluid spacetimes in Boyer--Lindquist coordinates. In the context of four-dimensional Einstein-Gauss-Bonnet gravity, the Boyer-Lindquist metric \cite{papnoi2022rotating,kumar2020rotating} describes the spacetime geometry of a rotating charged black hole. The metric is
\begin{equation}
\begin{aligned}
ds^{2}={}&
-\frac{\Delta}{\rho^{2}}
\left(dt-a\sin^{2}\theta\,d\phi\right)^{2}
+\frac{\rho^{2}}{\Delta}\,dr^{2}
+\rho^{2}\,d\theta^{2} \\
&+\frac{\sin^{2}\theta}{\rho^{2}}
\left[a\,dt-\left(r^{2}+a^{2}\right)d\phi\right]^{2},
\end{aligned}
\label{eq:metric}
\end{equation}

where
\begin{equation}
\begin{aligned}
\rho^{2} &= r^{2}+a^{2}\cos^{2}\theta,\\[2mm]
\Delta &= r^{2}+a^{2}
+\frac{r^{4}}{2\beta}
\left(
1-
\sqrt{
1+4\beta
\left(
\frac{2M}{r^{3}}
-\frac{Q^{2}}{r^{4}}
\right)
}
\right).
\end{aligned}
\label{eq:rho_delta}
\end{equation}
The rotation parameter, electric charge, and Gauss-Bonnet coupling parameter of the black hole are denoted by the parameters $a$, $Q$, and $\beta$ in the metric above. While the Gauss--Bonnet coupling is represented by the symbol $\alpha$ in the original formulation, we utilize $\beta$ throughout this work to clearly distinguish it from the spin parameter $a$. Additionally, only positive values of $\beta$ are taken into account. Notably, the solution simplifies to the well-known Kerr--Newman black hole in the limit $\beta \to 0$. 

The rotating EGBBH spacetime examined in this work is now presented. The following is an expression for the relevant spacetime metric in \cite{nosirov2024particle} is

\begin{align}
  ds^2=g_{tt}dt^2+g_{rr}dr^2 + g_{\theta\theta}\, d\theta ^ 2 
+ g_{\phi\phi}\, d\phi ^ 2 + 2g_{t\phi}\, dt \, d\phi,\label{e1} 
\end{align}
with the metric components provided by 
\begin{equation}
\begin{aligned}
g_{tt} &= -\frac{\Delta-a^{2}\sin^{2}\theta}{\rho^{2}}, \\
g_{rr} &= \frac{\rho^{2}}{\Delta}, \\
g_{\theta\theta} &= \rho^{2}, \\
g_{\phi\phi} &= \frac{\left(\left(r^{2}+a^{2}\right)^{2}
-\Delta a^{2}\sin^{2}\theta\right)\sin^{2}\theta}{\rho^{2}}, \\
g_{t\phi} &= \frac{a\sin^{2}\theta
\left(\Delta-\left(r^{2}+a^{2}\right)\right)}{\rho^{2}}.
\end{aligned}
\end{equation}

The black hole spacetime's electromagnetic four-potential can be written as
\begin{equation}
A_{\mu} dx^{\mu} =
\frac{Qr}{\rho^{2}}
\left(
dt - a\sin^{2}\theta  d\phi
\right).
\label{e2}
\end{equation}
For the minimally coupled Einstein--Maxwell--Gauss--Bonnet action~(1), the Maxwell
equations $\nabla_\mu F^{\mu\nu}=0$ do not involve the Gauss--Bonnet term, so
$\beta$ enters only through the metric function $\Delta(r)$. For the metric~(3)
with an arbitrary function $\Delta(r)$ one finds $\sqrt{-g}=\rho^2\sin\theta$, and
the potential~(7) satisfies
\begin{equation}
  \partial_\mu\!\left(\sqrt{-g}\,F^{\mu\nu}\right)=0
  \label{eq:maxwell}
\end{equation}
identically. Eq.~(7) is therefore an exact solution of the Maxwell equations on
the rotating EGB background, independent of $\beta$ to all orders. In the static limit $a=0$ it reduces to
$A_t=Q/r$, the known 4D EGB--Maxwell result.
The Cauchy horizon $r_{-}$ and the event horizon $r_{+}$ are the solutions to the constraint $\Delta = 0$, which defines the black hole horizons,
The corresponding surface gravity is determined by 

\begin{equation}
\kappa=\frac{\Delta'(r_{+})}{2\left(r_{+}^{2}+a^{2}\right)}\label{e3},
\end{equation}
where
\begin{eqnarray}
\Delta'(r_{+}) &=&
2r
-\frac{
\left(
\dfrac{4Q^{2}}{r^{5}}
-\dfrac{6M}{r^{4}}
\right)r^{4}
}{
\sqrt{
1+4\beta
\left(
-\dfrac{Q^{2}}{r^{4}}
+\dfrac{2M}{r^{3}}
\right)
}
}
\nonumber\\
&&
+\frac{
2r^{3}
\left[
1-
\sqrt{
1+4\beta
\left(
-\dfrac{Q^{2}}{r^{4}}
+\dfrac{2M}{r^{3}}
\right)
}
\right]
}{\beta}.
\end{eqnarray}

The frame-dragging effect is caused by the off-diagonal metric component $g_{t\phi}$ in rotating black hole geometries, which reflects the intrinsic link between the temporal and azimuthal coordinates. A co-rotating (frame-dragging) coordinate system produced by an appropriate coordinate transformation may be used to describe particle motion more conveniently. The equations governing geodesic trajectories, effective potentials, and orbital stability are simplified in this frame due to the local elimination of the mixed $dt,~d\phi$ component. Additionally, locally non-spinning observers, also known as Zero Angular Momentum Observers (ZAMOs), which are forced to co-rotate with the surrounding spacetime in the neighborhood of a rotating compact object, are naturally described by frame-dragging coordinates. The study of circular geodesics, innermost stable circular orbits (ISCOs), Lyapunov instability, chaotic particle dynamics, accretion disk structure, and energy-extraction processes like the Penrose process functioning inside the ergoregion all depend heavily on these coordinates. Frame-dragging coordinates offer a useful framework for characterizing rotating black hole spacetimes and comprehending the behavior of matter in strong gravitational fields from both a mathematical and scientific standpoint. For non-rotating spacetimes, they can be thought of as the natural equivalent of Schwarzschild coordinates \cite{Bardeen1972,Chandrasekhar1983,Misner1973}.

We use the following frame-dragging coordinate transformation \cite{chen2023chaos} to make the derivation of the Lyapunov exponent in  spinning spacetime easier,
\begin{align}
d\varphi& = d\phi - \omega\, dt  \notag \\
\rm {or} ~~~~~~~~d\phi  &= d\varphi + \omega\, dt ,\label{e5}
\end{align}
where 
\begin{align}
\omega =
\frac{a\left(\Delta-(r^{2}+a^{2}\right))}
{\left(r^{2}+a^{2}\right)^{2}-\Delta a^{2}\sin^{2}\theta}.
\end{align}
The spacetime line element is recast into the following form 
\begin{align*}
  ds^2 &= -F(r) \, dt^2 + \frac{1}{N(r)} \, dr^2 + C(r) \, d\theta^2 + D(r) \, d\varphi^2, 
  \end{align*}
\begin{align}
ds^2 &= -(\frac{\Delta-a^{2}\sin^{2}\theta}{\rho^{2}}
+\frac{a^{2}\sin^{2}\theta
\left(\Delta-\left(r^{2}+a^{2}\right)\right)^{2}}
{\rho^{2}\left(\left(r^{2}+a^{2}\right)^{2}
-\Delta a^{2}\sin^{2}\theta\right)})dt^2 \notag\\
&+\frac{\rho^2}{\Delta}\,dr^2
+\rho^2\,d\theta^2
+\frac{\left(\left(r^{2}+a^{2}\right)^{2}
-\Delta a^{2}\sin^{2}\theta\right)\sin^{2}\theta}
{\rho^{2}}\,d\varphi^2 .
\label{e6}
\end{align}
by combining the metric (\ref{e1}) with the electromagnetic potential of (\ref{e2}),
and the electromagnetic potential can be expressed as
\begin{align}
A_{\mu}dx^{\mu}
={}&
\frac{Qr}{\rho^{2}}
\left[
1
+
\frac{
a^{2}\sin^{2}\theta
\left(r^{2}+a^{2}-\Delta\right)
}{
\left(r^{2}+a^{2}\right)^{2}
-\Delta a^{2}\sin^{2}\theta
}
\right]dt
\nonumber\\
&-
\frac{aQr\sin^{2}\theta}{\rho^{2}}\,d\varphi.
\label{e7}
\end{align}
The above metric is very different from the standard form of a black hole spacetime with spherical symmetry. In the limit $a=0$, the electromagnetic four-potential maintains a nonvanishing $\phi$-component despite the geometry reducing to a spherically symmetric arrangement.

We note that the rotating metric~(3) is obtained from the static solution by the
Newman--Janis algorithm, and we do not claim that it solves the full
Einstein--Gauss--Bonnet field equations. Our analysis takes the metric~(3) and
the potential~(7) as the background on which the charged-particle dynamics is
studied.

\section{Hamiltonian formulation and Lyapunov exponent}\label{S3}

\subsection{Hamiltonian formulation}
The motion of a charged test particle (unit mass) in the rotating charged EGBBH background is described by
\begin{equation}
L=\frac{1}{2} g_{\mu\nu}\dot{x}^{\mu}\dot{x}^{\nu}-qA_{\mu}\dot{x}^{\mu}.\label{eq8}
\end{equation}
In this case, differentiation with regard to the correct time $\tau$ is indicated by $\dot{x}^{\mu}=dx^{\mu}/d\tau$, and the particle mass is set to unity. It is discovered that the expression derived from Eqs. ~(\ref{e6}) and ~(\ref{e7}) precisely coincides with the Lagrangian built from the electromagnetic potential~(\ref{e2}) and the spacetime metric~(\ref{e1}).

We use the latter formulation in the analysis that follows. Using the dragging-coordinate framework and limiting particle motion to the black hole's equatorial plane, denoted by $\theta=\pi/2$, the Lagrangian can be written as
\begin{align}
      L = \frac{1}{2} \left( -F \dot{t}^2 + \frac{\dot{r}^2}{N} + D \dot{\phi}^2 \right) - q A_t \dot{t} - q A_\phi \dot{\varphi},\label{e9}
\end{align}
where
\begin{align}
    F(r) & =\frac{\Delta-a^{2}}{\rho^{2}}  +\frac{a^{2}
\left(\Delta-\left(r^{2}+a^{2}\right)\right)^{2}}{\rho^{2}\left((r^{2}+a^{2}\right)^{2}-\Delta a^{2})}\notag \\
    N(r) & = \frac{\Delta}{\rho^2}, \quad 
    D(r) = \frac{\left(\left(r^{2}+a^{2}\right)^{2}
-\Delta a^{2}\right)}{\rho^{2}},\quad
    A_\phi = -\frac{aQr}{\rho^{2}}\notag\\
A_t & = \frac{Qr}{\rho^{2}}
\left(
1
+
a^{2}
\frac{(\Delta-(r^{2}+a^{2}))}{\left((r^{2}+a^{2}\right)^{2}-\Delta a^{2})}
\right).\label{e10}
\end{align}

The relevant momentum components are found in \cite{chen2023chaos} as 
\begin{align}
 \pi_t = -F \dot{t} - q A_t = E ,\quad 
  \pi_r = \frac{\dot{r}}{N},\quad 
  \pi_\phi = D \dot{\phi} - q A_\phi = L.\label{e11}
\end{align}
by using the generalized momentum definition $ \pi_\mu = \frac{\partial L}{\partial \dot{x}^\mu}$.
The particle's conserved energy and angular momentum are denoted by $E$ and $L$, respectively, in the aforementioned formulas. Next, the particle dynamics' controlling Hamiltonian is provided in \cite{chen2023chaos} by
\begin{align}
    H & =\frac{1}{2} (\pi_\mu + qA_\mu)(\pi_\nu + qA_\nu),\notag\\
    H &= \frac{-(\pi_t + qA_t)^2 +\pi_r^2 F N + (\pi_\phi+qA_\phi)^2D^{-1}F}{2 F}.\label{e12}
\end{align}
We derive the particle equations of motion within the Hamiltonian framework in order to study the system's chaotic behavior. The equations of motion that result are provided by \cite{chen2023chaos} 
\begin{align}
\begin{alignedat}{2}
\dot{t}
&= \frac{\partial H}{\partial\pi_t}
 = -\frac{\pi_t+qA_t}{F},
\qquad&
\dot{\pi}_t
&= -\frac{\partial H}{\partial t}=0,
\\[1mm]
\dot{r}
&= \frac{\partial H}{\partial\pi_r}
 = \pi_rN,
\qquad&
\dot{\phi}
&= \frac{\partial H}{\partial\pi_\phi}
 = \frac{\pi_\phi}{D},
\\[1mm]
\dot{\pi}_\phi
&= -\frac{\partial H}{\partial\phi}=0,
\qquad&
\dot{\pi}_r
&= -\frac{\partial H}{\partial r},
\end{alignedat}
\nonumber\\[1mm]
\dot{\pi}_r
= -\frac{1}{2}
\left[
\pi_r^2N'
-2qA_t'\frac{\pi_t+qA_t}{F}
+\frac{(\pi_t+qA_t)^2F'}{F^2}
\right.
\nonumber\\
\qquad\left.
-\frac{\pi_\phi^2D'}{D^2}
\right].
\label{e13}
\end{align}
In this case, differentiation with regard to the radial coordinate $r$ is indicated by the prime (${}^\prime$). The radial equation of motion at time $t$ takes the form \cite{Lei2021Chaos}:
\begin{align}
\frac{dr}{dt}
&= \frac{\dot{r}}{\dot{t}}
= -\frac{\pi_r F N}{\pi_t+qA_t},
\nonumber\\[2mm]
\frac{d\pi_r}{dt}
&= -qA_t'
+\frac{1}{2}
\left[
\frac{\pi_r^2 F N'}{\pi_t+qA_t}
+\frac{(\pi_t+qA_t)F'}{F}
\right.
\nonumber\\
&\qquad\left.
-\frac{
\left[(\pi_\phi+qA_\phi)^2D^{-1}\right]'F
}{
\pi_t+qA_t
}
\right].
\label{e14}
\end{align}
The four-velocity satisfies the normalization condition \(g_{\mu\nu}\dot{x}^\mu \dot{x}^\nu = \eta\) when the particle dynamics are restricted to the reduced phase space $(\pi_r,r)$, where $\eta=0$ describes null (massless) trajectories and $\eta=-1$ corresponds to timelike (massive) trajectories.

The normalizing condition results in the following constraint for a charged particle:
\[
\pi_t + q A_t = -\sqrt{F\left(1 + \pi_r^2 N + (\pi_\phi + q A_\phi)^2 D^{-1}\right)}.
\]
The phase-space variables are introduced. The particle dynamics equations, $F_{1}=\frac{dr}{dt}$ and $F_{2}=\frac{d\pi_{r}}{dt}$, can be rewritten as \cite{chen2023chaos}

\begin{align}
F_1
&=
\frac{\pi_r F N}
{\sqrt{
F\left[
1+\pi_r^2N
+\left(\pi_\phi+qA_\phi\right)^2D^{-1}
\right]
}},
\nonumber\\[2mm]
F_2
&=
-qA_t'
-\frac{
\pi_r^2(NF)'+F'
}{
2\sqrt{
F\left[
1+\pi_r^2N+\pi_\phi^2D^{-1}
\right]
}
}
\nonumber\\
&\quad
-\frac{
\left[
\left(\pi_\phi+qA_\phi\right)^2
D^{-1}F
\right]'
}{
2\sqrt{
F\left[
1+\pi_r^2N
+\left(\pi_\phi+qA_\phi\right)^2D^{-1}
\right]
}
}.
\label{e15}
\end{align}

\subsection{Separability of the Hamilton–Jacobi equation}
The inverse metric in Eq.~(\ref{eq:metric}) retains the Carter-separable structure for an arbitrary radial function $\Delta(r)$, as expressed in
\begin{equation}
\begin{split}
\rho^{2}g^{\mu\nu}\partial_{\mu}\partial_{\nu}
={}&\Delta\,\partial_{r}^{2}+\partial_{\theta}^{2}\\
&-\frac{1}{\Delta}\left[\left(r^{2}+a^{2}\right)\partial_{t}
+a\,\partial_{\phi}\right]^{2}\\
&+\frac{1}{\sin^{2}\theta}
\left[\partial_{\phi}+a\sin^{2}\theta\,\partial_{t}\right]^{2}.
\end{split}
\label{eq:inverse_metric}
\end{equation}
The electromagnetic potential given in Eq.~(\ref{e2}) is compatible with this separable structure through the identities
\begin{equation}
\begin{gathered}
\left(r^{2}+a^{2}\right)A_{t}+aA_{\phi}=Qr,\\
A_{\phi}+a\sin^{2}\theta\,A_{t}=0.
\end{gathered}
\label{eq:A_identities}
\end{equation}
Consequently, the charged-particle Hamilton--Jacobi equation admits an additive separation of variables. We introduce the Jacobi action
\begin{equation}
S=\frac{1}{2}\tau-Et+L\phi+S_{r}(r)+S_{\theta}(\theta),
\end{equation}
where $E$ and $L$ denote the conserved energy and azimuthal angular momentum, respectively. Substituting this ansatz into the Hamilton--Jacobi equation,
$\partial_{\tau}S+\mathcal{H}(x,\partial S)=0$, we obtain separate radial and polar equations,
\begin{align}
\Delta\left(\frac{dS_{r}}{dr}\right)^{2}
-\frac{\mathcal{P}^{2}}{\Delta}+r^{2}
&=-\mathcal{K},
\label{eq:HJ_r}\\
\left(\frac{dS_{\theta}}{d\theta}\right)^{2}
+\frac{\left(L-aE\sin^{2}\theta\right)^{2}}{\sin^{2}\theta}
+a^{2}\cos^{2}\theta
&=\mathcal{K},
\label{eq:HJ_theta}
\end{align}
where $\mathcal{K}$ is the separation constant and
\begin{equation}
\mathcal{P}(r)=E\left(r^{2}+a^{2}\right)-aL-qQr.
\label{eq:P}
\end{equation}

The corresponding first-order equations of motion are
\begin{align}
\rho^{2}\frac{dr}{d\tau}
&=\pm\sqrt{\mathcal{R}(r)},
\label{eq:rdot}\\
\rho^{2}\frac{d\theta}{d\tau}
&=\pm\sqrt{\Theta(\theta)},
\label{eq:thetadot}\\
\rho^{2}\frac{d\phi}{d\tau}
&=\frac{L}{\sin^{2}\theta}-aE
+\frac{a\,\mathcal{P}}{\Delta},
\label{eq:phidot}\\
\rho^{2}\frac{dt}{d\tau}
&=a\left(L-aE\sin^{2}\theta\right)
+\frac{\left(r^{2}+a^{2}\right)\mathcal{P}}{\Delta},
\label{eq:tdot}
\end{align}
with the radial and polar potentials given by
\begin{align}
\mathcal{R}(r)
&=\mathcal{P}^{2}-\Delta\left(r^{2}+\mathcal{K}\right),
\label{eq:R}\\
\Theta(\theta)
&=\mathcal{K}-a^{2}\cos^{2}\theta
-\frac{\left(L-aE\sin^{2}\theta\right)^{2}}
{\sin^{2}\theta}.
\label{eq:Theta}
\end{align}

The Carter constant can also be expressed in terms of the phase-space variables as
\begin{equation}
\begin{split}
\mathcal{K}={}&\pi_{\theta}^{2}
+\frac{\left(\pi_{\phi}
+a\sin^{2}\theta\,\pi_{t}\right)^{2}}{\sin^{2}\theta}
-2a^{2}\cos^{2}\theta\,\mathcal{H},
\end{split}
\label{eq:Carter}
\end{equation}
and satisfies
\begin{equation}
\left\{\mathcal{K},\mathcal{H}\right\}=0,
\qquad
\mathcal{K}=\mathcal{Q}+\left(L-aE\right)^{2},
\label{eq:Carter_props}
\end{equation}
where $\mathcal{Q}$ denotes Carter's constant in its usual normalisation. Thus, $\mathcal{K}$ is conserved along the particle trajectories.

Since the parameter $\beta$ enters the metric only through $\Delta(r)$, it does not affect the separability structure. The Hamilton--Jacobi equation therefore remains separable for arbitrary admissible values of $\beta$. In the limit $\beta\to0$, the equations reduce to those describing a charged particle in the Kerr--Newman spacetime. The existence of these constants establishes the Liouville integrability of the system. Specifically, $\mathcal{H}$, $E$, $L$, and $\mathcal{K}$ constitute four functionally independent integrals in involution for a system with four degrees of freedom. We verified the separability relations and the Poisson-bracket condition using computer algebra while retaining an arbitrary function $\Delta(r)$. Numerical integrations further confirm the conservation of $\mathcal{K}$, with relative deviations of approximately $10^{-9}$ along the trajectories. These results establish the integrability of the charged-particle dynamics in the absence of additional nonseparable interactions.

\subsection{Poincar\'e sections near the ISCO}
\label{sec:poincare_isco}

The phase-space structure near the innermost stable circular orbit (ISCO) can be
examined further with Poincar\'e sections. Since $t$ and $\phi$ are cyclic
coordinates, fixing the conserved quantities $E$ and $L$ reduces the dynamics to two
degrees of freedom, described by $(r,\theta,p_r,p_\theta)$, with
$p_r=(\rho^2/\Delta)\,dr/d\tau$ and $p_\theta=\rho^2\,d\theta/d\tau$. On the timelike
mass shell the Hamiltonian then takes the form
\begin{equation}
\mathcal{H}=\frac{1}{2\rho^{2}}\left[\Delta\,p_{r}^{2}+p_{\theta}^{2}
-\frac{\mathcal{P}^{2}(r)}{\Delta}
+\frac{\left(L-aE\sin^{2}\theta\right)^{2}}{\sin^{2}\theta}\right]=-\frac{1}{2},
\label{eq:H_reduced}
\end{equation}
where
\begin{equation}
\mathcal{P}(r)=E\left(r^{2}+a^{2}\right)-aL-qQr ,
\label{eq:P_isco_poincare}
\end{equation}

\begin{table*}[t]
\centering
\caption{Outer horizon, prograde ISCO of a particle with specific charge $q=0.5$,
and the orbital constants used for the near-ISCO sections of
Fig.~\ref{fig:poincare_isco}, chosen with Eq.~(\ref{eq:recipe}); $qQ>0$ is repulsive. Units $M=1$.}
\label{tab:isco}
\begin{tabular}{ccc|cccc|cc}
\hline
$a$ & $Q$ & $\beta$ & $r_{+}$ & $r_{\rm ISCO}$ & $E_{\rm ISCO}$ & $L_{\rm ISCO}$ & $E$ & $L$ \\
\hline
0.00 & 0.2 & 0.20 & 1.872 & 5.806 & 0.94701 & 3.2587 & 0.95084 & 3.3076 \\
0.35 & 0.2 & 0.20 & 1.789 & 4.494 & 0.93139 & 2.8836 & 0.93612 & 2.9268 \\
0.70 & 0.2 & 0.20 & 1.396 & 2.523 & 0.88159 & 2.2507 & 0.88891 & 2.2845 \\
\hline
0.30 & 0.0 & 0.20 & 1.836 & 4.780 & 0.92846 & 3.1180 & 0.93335 & 3.1648 \\
0.30 & 0.3 & 0.20 & 1.780 & 4.598 & 0.93669 & 2.8371 & 0.94112 & 2.8796 \\
0.30 & 0.6 & 0.20 & 1.579 & 3.958 & 0.93995 & 2.4071 & 0.94415 & 2.4432 \\
\hline
0.30 & 0.2 & 0.00 & 1.933 & 4.914 & 0.93660 & 2.9818 & 0.94096 & 3.0266 \\
0.30 & 0.2 & 0.20 & 1.812 & 4.698 & 0.93438 & 2.9446 & 0.93894 & 2.9888 \\
0.30 & 0.2 & 0.40 & 1.666 & 4.440 & 0.93158 & 2.9011 & 0.93641 & 2.9446 \\
\hline
\end{tabular}
\end{table*}

\begin{figure*}
\centering
\includegraphics[width=\hsize]{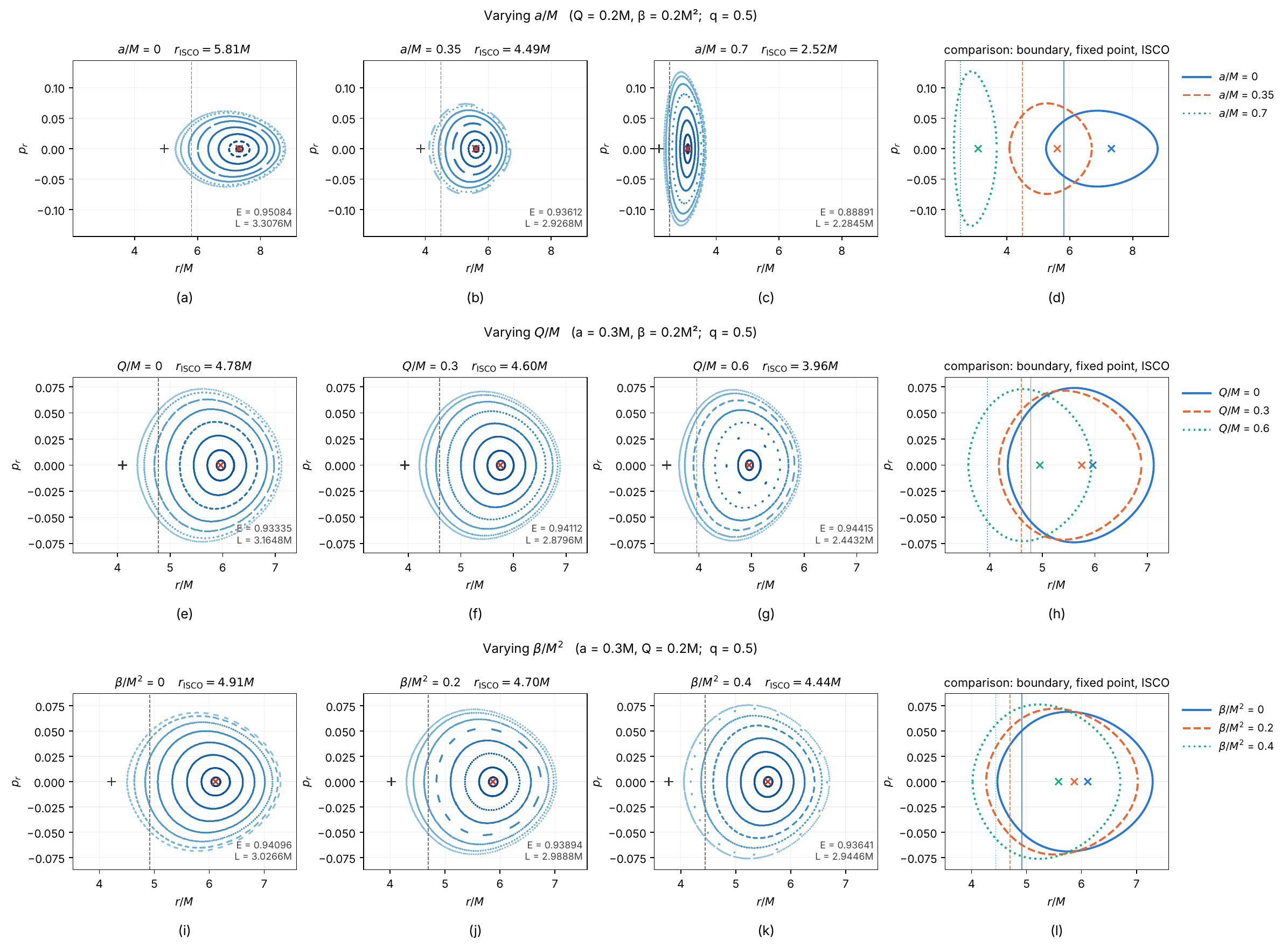}
\caption{Poincar\'e sections at $\theta=\pi/2$ ($\dot\theta>0$) of the integrable
motion of a charged particle near the prograde ISCO, for $q=0.5$ (repulsive for $Q>0$)
and $(E,L)$ chosen by Eq.~(\ref{eq:recipe}) (values in Table~\ref{tab:isco}).
(a) - (c): $a/M=0$, $0.35$, $0.7$ with $Q=0.2M$, $\beta=0.2M^2$;
(e) - (g): $Q/M=0$, $0.3$, $0.6$ with $a=0.3M$, $\beta=0.2M^2$;
(i) - (k): $\beta/M^2=0$, $0.2$, $0.4$ with $a=0.3M$, $Q=0.2M$.
Panels (d), (h) and (l) overlay the section boundaries and fixed points for the three
values of each row. Dashed lines mark $r_{\rm ISCO}$, $\times$ the fixed point and $+$
the unstable circular orbit.}
\label{fig:poincare_isco}
\end{figure*}

and Hamilton's equations give, after use of $\mathcal{H}=-1/2$,
\begin{align}
\frac{dr}{d\tau}&=\frac{\Delta\,p_{r}}{\rho^{2}},\qquad
\frac{d\theta}{d\tau}=\frac{p_{\theta}}{\rho^{2}},
\label{eq:eom_rtheta}\\
\frac{dp_{r}}{d\tau}&=\frac{1}{2\rho^{2}}\left[\frac{d}{dr}\!\left(\frac{\mathcal{P}^{2}}{\Delta}\right)
-\Delta'\,p_{r}^{2}-2r\right],
\label{eq:eom_pr}\\
\frac{dp_{\theta}}{d\tau}&=\frac{\sin\theta\cos\theta}{\rho^{2}}
\left[\frac{L^{2}}{\sin^{4}\theta}-a^{2}\left(E^{2}-1\right)\right],
\label{eq:eom_pth}
\end{align}
where $\Delta'=d\Delta/dr$. The particle charge enters only through $\mathcal{P}$, so
the polar equation is the same as for a neutral particle. We construct the Poincar\'e
section on the equatorial plane,
\begin{equation}
\theta=\frac{\pi}{2},\qquad \dot{\theta}>0,
\label{eq:section_isco}
\end{equation}
and represent the intersections in the $(r,p_r)$ plane.

Because the motion is integrable, every orbit lies on an invariant curve fixed by its
Carter constant $\mathcal{K}$. Using the radial first-order equation, the momentum on
the section can be written as
\begin{equation}
p_r=\pm\frac{\sqrt{\mathcal{R}(r)}}{\Delta(r)},\qquad
\mathcal{R}(r)=\mathcal{P}^{2}(r)-\Delta(r)\left(r^{2}+\mathcal{K}\right).
\label{eq:pr_poincare_isco}
\end{equation}
At $\theta=\pi/2$ the Carter constant is $\mathcal{K}=p_\theta^2+(L-aE)^2$, so it is
bounded below by its value for equatorial motion,
\begin{equation}
\mathcal{K}_{0}=\left(L-aE\right)^{2}\le\mathcal{K}\le\mathcal{K}_{c},\;\;\;
\mathcal{K}_{c}=\max_{r}\mathcal{K}(r),\;\;\;
\mathcal{K}(r)=\frac{\mathcal{P}^{2}(r)}{\Delta(r)}-r^{2},
\label{eq:K_range_isco}
\end{equation}
where $\mathcal{K}(r)$ is the Carter constant of the orbit that crosses the section at
radius $r$ with $p_r=0$. The equatorial orbit, $\mathcal{K}=\mathcal{K}_0$, forms the
boundary of the section, and the fixed point at $r=r_c$, where $\mathcal{K}(r)$
reaches $\mathcal{K}_c$, is a spherical orbit of constant radius.

The equatorial effective potential is obtained by solving $\mathcal{R}(r)=0$ with
$\mathcal{K}=\mathcal{K}_0$ for the energy; the root corresponding to
future-directed motion is
\begin{equation}
E=V(r;L)=
\frac{-\mathcal{B}+\sqrt{\mathcal{B}^{2}-\mathcal{A}\mathcal{C}}}
{\mathcal{A}},
\label{eq:Veff_isco_poincare}
\end{equation}
where
\begin{align}
\mathcal{A}&=(r^2+a^2)^2-a^2\Delta,\nonumber\\
\mathcal{B}&=-(r^2+a^2)(qQr+aL)+aL\Delta,\label{eq:ABC_isco_poincare}\\
\mathcal{C}&=(qQr+aL)^2-\Delta(r^2+L^2).\nonumber
\end{align}
Circular orbits are determined by
\begin{equation}
\frac{\partial V}{\partial r}=0,
\end{equation}
whereas the ISCO is identified by the additional marginal-stability condition
\begin{equation}
\frac{\partial V}{\partial r}
=\frac{\partial^2 V}{\partial r^2}=0.
\label{eq:isco_marginal_poincare}
\end{equation}

For each black hole we first compute the prograde ISCO of the charged particle, by
bisection in $L$ for the appearance of a pair of circular orbits followed by a Newton
solution of Eq.~(\ref{eq:isco_marginal_poincare}). The orbital constants are then
chosen as
\begin{equation}
L=\left(1+\epsilon\right)L_{\rm ISCO},\qquad
E=E_{\min}+f\left(E_{\max}-E_{\min}\right),
\label{eq:recipe}
\end{equation}
with $\epsilon=0.015$ and $f=0.9$, where $E_{\min}=V(r_s;L)$ and $E_{\max}=V(r_u;L)$
are the values of Eq.~(\ref{eq:Veff_isco_poincare}) at the stable ($r_s$) and
unstable ($r_u$) circular orbits for this $L$. All cases are therefore equally close
to their own ISCO, and $f<1$ keeps the outermost orbits away from the separatrix
through $r_u$. Each orbit starts on the section with
\begin{equation}
\begin{gathered}
\theta_{0}=\frac{\pi}{2},\qquad p_{r0}=0,\qquad
r_{0}=r_{\rm in}+f_{k}\left(r_{c}-r_{\rm in}\right),\\
p_{\theta0}=\sqrt{\mathcal{K}(r_{0})-\mathcal{K}_{0}},
\end{gathered}
\label{eq:ic_isco_poincare}
\end{equation}
where $r_{\rm in}$ is the inner radius at which $\mathcal{K}(r)=\mathcal{K}_0$,
$f_k=0.02$, $0.10$, $0.22$, $0.36$, $0.52$, $0.68$, $0.84$ and $0.95$, and the
positive root gives $\dot\theta>0$.

Equations~(\ref{eq:eom_rtheta}) - (\ref{eq:eom_pth}) were integrated in
Boyer--Lindquist coordinates with an explicit Runge--Kutta 8(5,3) scheme (DOP853),
with relative and absolute tolerances of $10^{-11}$; the crossings of the
section~(\ref{eq:section_isco}) were located by event detection, with 300 crossings
per orbit and eight orbits per panel. Throughout, $\Delta$ is evaluated in the form
\begin{equation}
\Delta=r^{2}+a^{2}-\frac{2\left(2Mr-Q^{2}\right)}
{1+\sqrt{1+4\beta\left(2M/r^{3}-Q^{2}/r^{4}\right)}} ,
\label{eq:Delta_stable}
\end{equation}
which is algebraically identical to the metric function of Sec.~\ref{s2} but free of
round-off as $\beta\to0$. Over all 72 orbits, $|\mathcal{H}+1/2|\le3\times10^{-11}$,
no orbit approached the horizon, and every crossing lies on its analytic
curve~(\ref{eq:pr_poincare_isco}) with $\mathcal{K}=\mathcal{K}(r_0)$ to within
$7\times10^{-7}$ in $p_r$.

Figure~\ref{fig:poincare_isco} shows Poincar\'e sections of the charged-particle
motion near the prograde ISCO for the three parameter sequences of
Table~\ref{tab:isco}, and so gives a phase-space picture of the approach to marginal
stability. We use the same particle charge, $q=0.5$, as in the trap study of
Sec.~\ref{S4}, and choose $(E,L)$ with Eq.~(\ref{eq:recipe}), so that the orbits are
bound in the shallow potential well that exists just above $L_{\rm ISCO}$, close to
the black hole, where the influence of $\beta$ is largest.

In every panel the orbits lie on smooth closed curves surrounding a single fixed
point, with no island chains and no scattered points. The numerical points agree with
the analytic curves of Eq.~(\ref{eq:pr_poincare_isco}). Curves that appear dashed belong to
tori close to a resonance, sampled by a finite number of crossings; they are not
chains of islands. The motion is therefore regular for all values of $a$, $Q$ and
$\beta$ considered, and these sections are the integrable baseline against which the
trapped dynamics of Sec.~\ref{S4} is compared.

All three parameters shift the section towards the black hole together with the
ISCO. Increasing the spin from $a=0$ to $0.7M$ moves $r_{\rm ISCO}$ from $5.81M$ to
$2.52M$, increasing the charge from $Q=0$ to $0.6M$ moves it from $4.78M$ to $3.96M$,
and increasing the coupling from $\beta=0$ to $0.4M^2$ moves it from $4.91M$ to
$4.44M$. The spin also changes the shape of the section most strongly: at $a=0.7M$
the section is narrow in $r$ and extends to larger $|p_r|$ [panel (d)], whereas
varying $Q$ or $\beta$ mainly displaces sections of similar shape [panels (h) and
(l)]. The larger values of $|p_r|$ close to the hole partly reflect the relation
$p_r=(\rho^2/\Delta)\,dr/d\tau$, since $\Delta$ decreases towards the horizon, and
should not be read as a larger radial velocity alone.

The effect of $\beta$ is the weakest of the three. To first order in $\beta$,
Eq.~(\ref{eq:Delta_stable}) gives
\begin{equation}
\Delta\simeq r^{2}+a^{2}-2Mr+Q^{2}
+\frac{\beta}{r^{2}}\left(2M-\frac{Q^{2}}{r}\right)^{2},
\label{eq:Delta_beta_expansion}
\end{equation}
so the GB correction decays as $r^{-2}$, whereas the Kerr--Newman part grows as
$r^{2}$. It is therefore significant only close to the black hole, which is why we
examine sections near the ISCO. Since the correction is positive for $\beta>0$, it
can be read as a reduction of the effective mass at small $r$, consistent with the
inward shift of $r_+$ and $r_{\rm ISCO}$ as $\beta$ increases; this is an
interpretation of the trend rather than a separate result.

\subsection{Lyapunov exponent}
Using the matrix approach, the Lyapunov exponent is calculated by building the Jacobian matrix $K_{ij}$ in the phase space $(\pi_r,r)$. The Jacobian-matrix method used here is the general Lyapunov analysis used in the previous studies such as Ref.~\cite{chen2023chaos}. We stress that the novelty of the present paper is not the new definition of the Lyapunov exponent or the new Jacobian formalism. Instead, the method is specially tailored to the rotating charged Einstein-Gauss-Bonnet geometry. The equations of motion and equilibrium conditions of charged particles are changed by the Gauss--Bonnet coupling. Such changes cascade through the elements of the Jacobian matrix and result in a specific dependence of the Lyapunov exponent, as well as violation of chaos on the black hole charge, rotation parameter and the Gauss--Bonnet coupling. The following analysis is thus not just a simple replacement of a single metric function, but will also enable us to investigate the effect of the higher-curvature corrections on the stability of equilibrium orbits in a rotating black hole spacetime.

The authors in \cite{chen2023chaos} defines its components which are given by  \begin{align}
     K_{11} &= \frac{\partial F_1}{\partial r}\quad
       K_{12} = \frac{\partial F_1}{\partial \pi_r}\quad
        K_{21}= \frac{\partial F_2}{\partial r}\quad
         K_{22} = \frac{\partial F_2}{\partial \pi_r}.\label{e16}
 \end{align}
Finding the equilibrium point in the phase space, represented by $r_{0}$, is the first step in the analysis. The particle's radial momentum at this equilibrium point satisfies\cite{YuChenGao2022,WangChen2023}
 \begin{align}
     \pi_r = \frac{d\pi_r}{dt} =0.\label{e17}
 \end{align}
The preceding formula results in
\begin{align}
&-qA_t'
-\frac{
\pi_r^2(NF)'+F'
}{
2\sqrt{
F\left(
1+\pi_r^2N+
(\pi_\phi+qA_\phi)^2D^{-1}
\right)
}
}
\nonumber\\
&\quad
-\frac{
\left[
(\pi_\phi+qA_\phi)^2D^{-1}F
\right]'
}{
2\sqrt{
F\left(
1+\pi_r^2N+
(\pi_\phi+qA_\phi)^2D^{-1}
\right)
}
}
=0.
\label{e18}
\end{align}
The radial coordinates at which equilibrium orbits \cite{chen2023chaos} occur are specified by the equation above.
The Lyapunov exponent 
\begin{align}
\lambda^2
={}&
\frac{N
\left[
F'
+(\pi_\phi+qA_\phi)^2
(D^{-1}F)'
\right]^2
}{
4F
\left[
1+(\pi_\phi+qA_\phi)^2D^{-1}
\right]^2
}
\nonumber\\
&-
\frac{N}{2}
\frac{
F''
+
\left[
(\pi_\phi+qA_\phi)^2
D^{-1}F
\right]''
}{
1+(\pi_\phi+qA_\phi)^2D^{-1}
}
\nonumber\\
&-
\frac{
qA_t''FN
}{
\sqrt{
F\left[
1+(\pi_\phi+qA_\phi)^2D^{-1}
\right]
}
}.
\label{e19}
\end{align}
can be found by combining Eqs. ~(\ref{e16}) and ~(\ref{e17}).
At the event horizon, where $\Delta=0$, evaluating Eqs.~(\ref{e3}), (\ref{e10}), and~(\ref{e19}) produces
\begin{align}
      \lambda^2 &= \frac{1}{4} \frac{N(F')^2}{F}=\kappa^2.\label{e20}
\end{align}
The chaotic bound is saturated near the event horizon. This result is in line with earlier research on spherically symmetric black holes \cite{zhao2018static,lei2022circular}.

\begin{table*}[ht]
    \centering
  \caption{ Equilibrium point distribution for a charged particle in the spacetime of a non-extremal spinning EGBBH with fixed parameters $\beta=0.10$ and $a=0$. For $Q=0.40$, $0.50$, $0.60$, and $0.70$, the corresponding event horizon radii are $r_{+}=1.86023$, $1.806226$, $1.73485$, and $1.64031$.}
    \label{t1}
    \begin{tabular}{|c| c| c| c| c| c| c| c| c|}
        \toprule
        \hline
         L & 0 & 1  & 3 & 5 & 7& 10 & 20 \\
         \hline
        \midrule
        $Q=0.40$  & 1.95551 & 1.97584 & 2.087252 & 2.20309 & 2.29496 & 2.39484 & 2.56586 \\
        \hline
       $Q=0.50$  & 1.85833 & 1.87189 & 1.95569 & 2.05758 & 2.14691 & 2.25078 & 2.44119\\
        \hline
       $Q=0.60$  &1.76463 & 1.77365 & 1.83418 & 1.91785 & 1.9988 & 2.1001 & 2.30093\\
        \hline
      $Q=0.70$  & 1.65707 & 1.66298 & 1.70516 & 1.77011 & 1.8394 & 1.93344 & 2.13767\\
        \hline
        \bottomrule
    \end{tabular}
\end{table*}

\begin{table*}[ht]
    \centering
   \caption{Equilibrium point distribution for a charged particle in the spacetime of a non-extremal spinning EGBBH with fixed parameters $\beta=0.30$ and $Q=0.30$. For $a=0.30$, $0.40$, and $0.50$,  the corresponding event horizon radii are $r_{+}=1.70796$, $1.64295$, and $1.54183$.}
    \label{t2}
    \begin{tabular}{|c| c| c| c| c| c| c| c| c|}
    \hline
        \toprule
        $L$  & -3 & -2 & -1 & 0 & 1 & 2 & 3 \\
        \hline
            $a=0.30$  & 2.0389613 & 1.965454 & 1.893993 & 1.84686 & 1.854618 & 1.9205976 & 2.009239\\
        \hline
        $a=0.40$  &  1.986125 & 1.910097 & 1.834899 & 1.780683 & 1.77846 & 1.8430265 & 1.941713\\
        \hline
        $a=0.50$  &  1.90406 & 1.821418 & 1.738498 & 1.674460 & 1.658617 & 1.717904 & 1.834711\\
        \hline
           \end{tabular}
\end{table*}

\begin{table*}
    \centering
    \caption{Equilibrium point distribution for a charged particle around a non-extremal rotating EGBBH with fixed rotation and charge parameters, $a=Q=0.30$. For $\beta=0.10$, $0.20$, and $0.40$, the corresponding event horizon radii are $r_{+}=1.84537$, $1.78005$, and $1.62635$, respectively.}
    \label{t3}
    \begin{tabular}{|c| c| c| c| c |c |c |c| c|}
       
        \hline
      $L$  & -3 & -2 & -1 & 0 & 1 & 2 & 3 \\
        \hline
$\beta=0.10$  &  2.21981 & 2.154905 & 2.091241 &
2.050692 & 2.064638 & 2.12877 & 
2.20597\\
        \hline
        $\beta=0.20$  & 2.134250 & 2.065030 & 1.99701 & 1.952486 & 1.963708 & 2.030024 & 2.113276 \\
        \hline
        $\beta=0.40$  & 1.930027 & 1.852458 & 1.779253 & 1.731819 & 1.73561 & 1.797227 & 1.88928 \\
        \hline
       \end{tabular}
\end{table*}

\begin{table*}
    \centering
  \caption{Charged particle equilibrium orbits around an extremal spinning EGB black hole for $\beta =0.10$. For the first, second, and third rows, the corresponding event horizon radii are $r_{+}=1.01705$, $1.00335$, and $1.01532$.}
    \label{t5}
    \begin{tabular}{|c| c| c| c| c |c |c |c| c|}
        \toprule
        \hline
       $L$  & -3 & -2 & -1 & 0 & 1 & 2 & 3 \\
        \hline
        
        \midrule
      $Q =\sqrt{\frac{20}{25}}$,$a =\sqrt{\frac{21}{25}}$  & 2.222388 & 2.188760 & 2.154917 & 2.121794 & 2.090889 & 2.064574 & 2.046543 \\
        \hline
        $Q =\sqrt{\frac{22}{25}}$,$a =\sqrt{\frac{19}{25}}$  & 2.110118& 2.078920& 2.047669& 2.017173& 1.988668& 1.964033& 1.946092\\
        \hline
         $Q =\sqrt{\frac{24}{25}}$,$a =\sqrt{\frac{17}{25}}$  &1.988716& 1.960005& 1.931427& 1.903684& 1.877802& 1.855266& 1.838208\\
         \hline
       
    \end{tabular}
\end{table*} 
\begin{table*}
    \centering
\caption{Charged particle equilibrium orbits around an extremal spinning EGB black hole for $Q =0.60$. For the first, second, and third rows, the corresponding event horizon radii are $r_{+}=1.03695$, $1.06646$, and $1.09254$.}
    \label{t7}
    \begin{tabular}{|c| c| c| c| c |c |c |c| c|}
        \toprule
        \hline
       $L$  & -3 & -2 & -1 & 0 & 1 & 2 & 3 \\
        \hline
        \midrule 
      $a =\sqrt{\frac{24}{25}}$,$\beta =\sqrt{\frac{8}{25}}$  & 2.463983 & 2.416457 & 2.368575 & 2.323340 & 2.286379 & 2.267597& 2.281265 \\
        \hline
        $a =\sqrt{\frac{22}{25}}$,$\beta =\sqrt{\frac{6}{25}}$  & 2.431166 &2.383830 & 2.336174 & 2.291293 & 2.254988 & 2.237395 & 2.252605 \\
        \hline
         $a =\sqrt{\frac{20}{25}}$,$\beta =\sqrt{\frac{4}{25}}$  & 2.404470 & 2.357377 & 2.310015 & 2.265603 & 2.230194 & 2.214266 & 2.231652 \\
          \hline
        \bottomrule
    \end{tabular}
\end{table*}

\begin{figure}
\centering
\includegraphics[width=0.7\linewidth,height=2.8in]{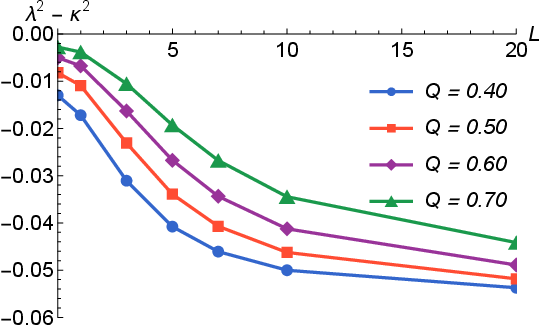}
\caption{ In this figure,
angular momentum $L$ is represented by the horizontal line, and $\lambda^2-\kappa^2$ is represented by the vertical line.
The graphic shows how $\lambda^2-\kappa^2$ varies with angular momentum $L$ for various values of $Q$. $\lambda^2-\kappa^2$ declines and becomes increasingly negative as $L$ increases, suggesting improved stability. Conversely, increasing $Q$ causes the curves to migrate higher, which lessens the negative value of $\lambda^2-\kappa^2$ and weakens the system's stability.
}
\label{f1}
\end{figure}

\begin{figure}
\centering
\includegraphics[width=0.7\linewidth,height=2.8in]{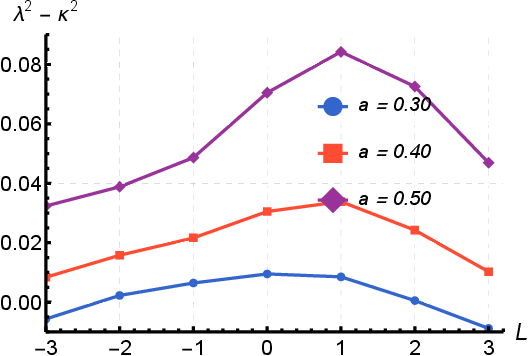}
\caption{
The fluctuation of $\lambda^{2}-\kappa^{2}$ with respect to $L$ for various values of the parameter $a$ ($a=0.30$, $0.40$, and $0.50$) is shown in this figure. $\lambda^2-\kappa^2$ increases with increasing $L$, reaching a peak around $L=1$, after which it progressively decreases. Throughout the spectrum under consideration, positive values of  $\lambda^2-\kappa^2$ indicate unstable motion, and bigger values of $a$ increase the severity of this instability.}
\label{f2}
\end{figure}
\begin{figure}
\centering
\includegraphics[width=0.7\linewidth,height=2.8in]{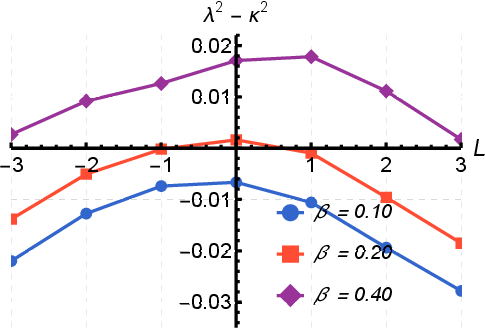}
\caption{
This figure shows how $\lambda^2-\kappa^2$ varies with angular momentum $L$ for various $\beta$ values. $\lambda^2-\kappa^2$ is primarily negative for $\beta=0.10$ and $0.20$, showing steady motion. On the other hand, it stays positive for $\beta=0.40$, suggesting unstable motion. In general, the curves rise toward $L\approx1$, where the maximum value is reached, and then fall as $L$ grows. The curves shift higher as $\beta$ increases, increasing $\lambda^2-\kappa^2$ and resulting in a change from stable to unstable behavior.
}
\label{f3}
\end{figure}

\begin{figure}
\centering
\includegraphics[width=0.7\linewidth,height=2.8in]{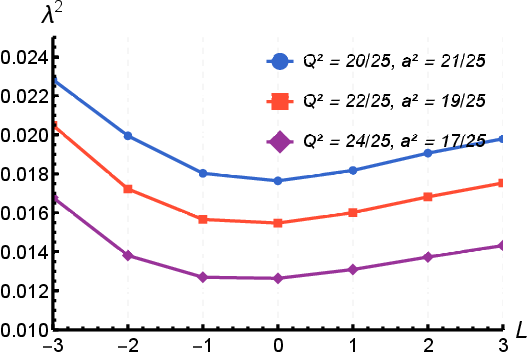}
\caption{
For an extremal rotating EGBBH with a fixed value of $\beta=0.10$, the Lyapunov exponent $\lambda^{2}$ depends on the particle angular momentum $L$. The picture illustrates how angular momentum affects how  particle motion is around a black hole.
}
\label{f5}
\end{figure}

\begin{figure}
\centering
\includegraphics[width=0.7\linewidth,height=2.8in]{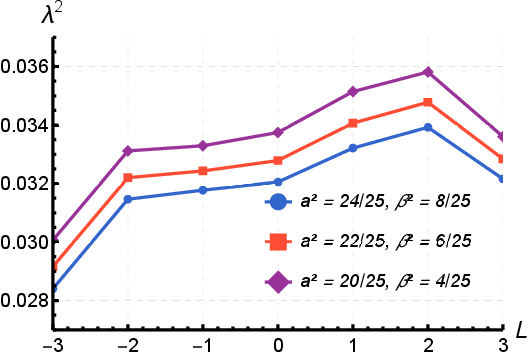}
\caption{
Lyapunov exponent $\lambda^{2}$ behavior for an extremal rotating EGBBH at a constant value of $Q =0.60$ as a function of particle angular momentum $L$. The graphic highlights how angular momentum affects test particle orbital stability.
}
\label{f7}
\end{figure}

\paragraph*{Meaning and range of the bound.} The Maldacena--Shenker--Stanford bound $\lambda\le2\pi T/\hbar$ is a statement about out-of-time-order correlators of a quantum system at temperature $T$. For a black hole $T=\kappa/2\pi$, which gives the form $\lambda\le\kappa$ used throughout this literature. Applying it to the classical Lyapunov exponent of a probe particle is a heuristic extension. Eq.~(\ref{e20}) shows that the bound is saturated when the equilibrium orbit approaches the horizon, and the question studied below is how $\lambda^2-\kappa^2$ behaves for equilibrium radii $r_0>r_+$. Two consequences follow. First, the bound is a statement at $T>0$: for $\kappa=0$ it reads $\lambda^2\le0$ and cannot be tested in a meaningful way (Sec.~\ref{s3}). Second, a positive $\lambda^2-\kappa^2$ at finite $r_0$ should be read as a property of the probe-particle model on the chosen background, subject to the caveats in Sec.~\ref{s5}.

\section{Chaos Bound and Its Violation for Rotating Einstein-Gauss-Bonnet Black Holes} \label{s3}
The dynamics of a charged particle in the spacetime of a spinning charged EGBBH is examined in order to determine the Lyapunov exponent of chaotic motion. The balance between the Lorentz and centrifugal forces determines the particle's equilibrium, and appropriate adjustments to the particle's charge and angular momentum can govern its radial location. In the near-horizon region of spherically symmetric black holes, previous research \cite{zhao2018static,lei2022circular} examined violations of the chaos bound. In this analysis, we fix the particle charge and investigate the effect of changes in the angular momentum on the Lyapunov exponent. The range of angular momentum and the spatial locations where the chaos bound is breached are identified with special care. The related equilibrium locations are anticipated to change in accordance with the explicit dependence of the equilibrium configurations on the angular momentum of the particle. 
The angular momentum of the particle has a sensitive effect on the equilibrium radius. Consequently, Eqs.~(\ref{e15}) and~(\ref{e17}) are solved numerically to get the equilibrium configurations. Tables~\ref{t1}–\ref{t7} list the corresponding equilibrium radii and the related event horizon radius for each set of parameters. The particle charge and black hole mass are kept at $q=10$ and $M=1$, respectively, throughout the investigation.

The direction of the particle's travel in relation to the rotation of the black hole is another important component of its dynamics. The particle co-rotates with the black hole, hence its angular momentum will be positive, otherwise the particle counter-rotates with the black hole, hence its angular momentum is negative. The numerical solutions shown in the figure show that with increasing rotational momentum of the particle, the equilibrium radius increases. Table~\ref{t1}. The result indicates that there is a positive correlation in the two parameters of rotational momentum and the equilibrium radius, i.e. larger rotational momentum implies greater equilibrium radius.

A numerical study of the Lyapunov exponent and the surface gravity based on Eqs.~(\ref{e3}), (\ref{e10}), (\ref{e15}), (\ref{e17}), and~(\ref{e19}) is used to characterize the chaotic dynamics. We calculate the amount $\lambda^{2}-\kappa^{2}$ in order to verify the validity of the chaotic bound. According to Maldacena \textit{et al.}~\cite{Maldacena:2015waa}, a negative value of this number, $\lambda^{2}-\kappa^{2}<0$, indicates that the chaos bound is satisfied.
Conversely, a positive value of $\lambda^{2}-\kappa^{2}$ denotes a violation of the chaotic bound.

The behaivour of $\lambda^{2}-\kappa^{2}$ and angular momentum $L$ is displayed in \textbf{Figs.}~\ref{f1}--\ref{f3}. In \textbf{Fig.}~\ref{f1}, we analyze that, how $\lambda^2-\kappa^2$ varies with angular momentum $L$ for various values of $Q$. $\lambda^2-\kappa^2$ declines and becomes increasingly negative as $L$ increases, suggesting improved stability. Conversely, increasing $Q$ causes the curves to migrate higher, which lessens the negative value of $\lambda^2-\kappa^2$ and weakens the system's stability.

The variation of $\lambda^{2}-\kappa^{2}$ with the angular momentum $L$ for various values of the rotation parameter $a$ is displayed in \textbf{Fig.}~\ref{f2}. In this figure, $\lambda^2-\kappa^2$ increases with increasing $L$, reaching a peak around $L=1$, after which it progressively decreases. Throughout the spectrum under consideration, positive values of  $\lambda^2-\kappa^2$ indicate unstable motion, and bigger values of $a$ increase the severity of this instability.

The dependency of $\lambda^{2}-\kappa^{2}$ on the angular momentum $L$ for various values of the parameter $\beta$ is shown in \textbf{Fig.}~\ref{f3}. $\lambda^2-\kappa^2$ is primarily negative for $\beta=0.10$ and $0.20$, showing steady motion. On the other hand, it stays positive for $\beta=0.40$, suggesting unstable motion. In general, the curves rise toward $L\approx1$, where the maximum value is reached, and then fall as $L$ grows. The curves shift higher as $\beta$ increases, increasing $\lambda^2-\kappa^2$ and resulting in a change from stable to unstable behavior.

The numerical analysis also shows how the black hole parameters affect the surface gravity and the event horizon radius. One parameter is changed at a time in the current study while the other values remain constant. We discover that both the event horizon radius and the surface gravity drop when any of the factors under consideration are increased. Additionally, our current findings are different from those presented in \cite{chen2023chaos}, where the event horizon radius increased as each of the characteristics taken into consideration increased. The current analysis, on the other hand, clearly differs qualitatively from the results of \cite{chen2023chaos}  by demonstrating that increasing any of the parameters taken into consideration consistently reduces both the event horizon radius and the surface gravity. 

 \subsection*{Extremal configurations as an instability diagnostic}
The inner and outer horizons coincide in the extremal limit, so $\kappa=0$ and $T=0$. In this case the $\kappa$-form of the bound reads $\lambda^2\le0$, and every unstable equilibrium orbit has $\lambda^2>0$ by construction. A positive $\lambda^2$ is therefore \emph{not} an independent violation of the chaos bound, because the derivation of the bound requires a non-zero horizon temperature. What the extremal case does provide is (i) the continuous $\kappa\to0$ limit of the non-extremal quantity $\lambda^2-\kappa^2$, and (ii) a clean measure of how unstable the orbits are, through the magnitude of $\lambda^2$ and the radial interval over which $\lambda^2>0$. We report these quantities in Figs.~\ref{f5}--\ref{f7}. Since the Lyapunov exponent is evaluated at $r_0>r_+$, it stays finite as $\kappa\to0$, and the extremal curves should be read as an upper reference for the near-extremal non-extremal results.

Fig.~\ref{f5} (fixed $\beta=0.10$) and Fig.~\ref{f7} (fixed $Q=0.60$) show $\lambda^2$ against $L$. In Fig.~\ref{f5}, for $Q=\sqrt{\frac{20}{25}}$ and $a=\sqrt{\frac{21}{25}}$, the chaos bound is violated within the radial interval, which belongs to  $[2.1549,\,2.222]$, when the particle and black hole counter-rotate, while the corresponding radial interval belongs to  $[2.0465,\,2.1218]$ when they co-rotate.
For $Q = \sqrt{\frac{22}{25}}$ and $a=\sqrt{\frac{19}{25}}$, the chaos bound is violated within the radial interval, which belongs to  $[2.0477,\,2.1101]$, when the particle and black hole counter-rotate, while the corresponding radial interval belongs to  $[1.9461,\,2.0172]$ when they co-rotate. For $Q = \sqrt{\frac{24}{25}}$ and $a = \sqrt{\frac{17}{25}}$, a similar pattern emerges, with corresponding intervals belonging to $[1.9314,\,1.9887]$ in the counter-rotating case and $[1.8382,\,1.9037]$ in the co-rotating configuration.
In \textbf{Fig.}~\ref{f7}, the impact of the parameter $a$ and the parameter $\beta$ is further demonstrated. For $a = \sqrt{\frac{24}{25}}$ and $\beta = \sqrt{\frac{8}{25}}$, the chaos bound is violated within the radial interval, which belongs to  $[2.3686,\,2.4639]$, when the particle and black hole counter-rotate, while the corresponding radial interval belongs to  $[2.2813,\,2.3233]$, when they co-rotate. When the parameters are chosen as $a = \sqrt{\frac{22}{25}}$ and $\beta = \sqrt{\frac{6}{25}}$, the chaos bound is violated within the radial interval, which belongs to  $[2.3362,\,2.4312]$, when the particle and black hole counter-rotate, while the corresponding radial interval belongs to  $[2.2526,\,2.2913]$, when they co-rotate. The corresponding radical regions in the case of $a= \sqrt{\frac{20}{25}}$ and $\beta = \sqrt{\frac{4}{25}}$ belong to $[2.3100,\,2.4045]$ for counter-rotating trajectories and $[2.2317,\,2.2656]$ for co-rotating trajectories.  
These results show that, in the extremal case, the charge, spin and Gauss--Bonnet coupling shift the equilibrium radii and the width of the radial interval in which $\lambda^2>0$. In the data above the unstable interval is wider for co-rotating orbits in Fig.~\ref{f5} (varying $Q$ and $a$) and wider for counter-rotating orbits in Fig.~\ref{f7} (varying $a$ and $\beta$), so no single ordering holds. For non-extremal black holes the sign of $\lambda^2-\kappa^2$ is fixed by the competition among the three terms of Eq.~(\ref{e19}): the first two terms come from the metric functions and their derivatives and the last one is the electric-force term $-qA_t''FN/\sqrt{\cdots}$. Increasing $Q$, $a$ or $\beta$ raises $\lambda^2-\kappa^2$ in the sampled range (Figs.~\ref{f1}--\ref{f3}), while increasing $|L|$ beyond $|L|\sim3$ lowers it, so that large-$|L|$ orbits are on the stable side.

The behavior of \textbf{Figs.}~\ref{f1}--\ref{f7} shows that the validity of the sign of $\lambda^2-\kappa^2$ for rotating EGBBHs depends on the interplay of the electric charge $Q$ and the rotation parameter $a$ and Gauss-Bonnet Coupling $\beta$ as well as the particle angular momentum $L$. The results confirm that the rotational and nonlinear electromagnetic effects are key to the stable/chaotic particle motion regime transition near the black hole horizon.

The extremal configurations shown in \textbf{Figs.}~\ref{f5}--\ref{f7} remain unstable ($\lambda^2>0$) over the whole range of $L$ explored, which is expected at $\kappa=0$ and, as explained above, is not by itself a violation of the bound. In In \textbf{Fig.}~\ref{f5}, the unstable interval is wider for co-rotating than for counter-rotating orbits for all three $(Q,a)$ pairs, whereas in \textbf{Fig.}~\ref{f7} it is wider for counter-rotating orbits for all three $(a,\beta)$ pairs; the rotational asymmetry therefore depends on which parameters are varied.
Finally, as seen in the Fig.~\ref{f7}, both the electric charge and the GB parameter also affect the radial width of the unstable regions. The figure clearly illustrates the regulation of chaotic motion onset and persistence by both gravitational attraction and black hole rotation, and by the nonlinear electromagnetic interactions. The chaotic dynamics of charged particles around rotating EGBBHs described in the analysis of \textbf{Figs.}~\ref{f1}--\ref{f7} is caused by the delicate balance between these competing physical effects.

\section{Chaotic dynamics of a trapped charged particle near the rotating EGB black hole}\label{S4}
The charged-particle dynamics discussed in Sec.~\ref{s2} is integrable owing to the existence of the Carter constant. Consequently, the onset of chaotic behaviour requires an additional perturbation that breaks the separability of the Hamilton--Jacobi equation. Following the approach adopted in near-horizon chaos studies, we introduce an external harmonic potential that confines the charged particle near the black hole horizon. Such a potential provides a phenomenological description of a particle subjected to external restoring forces. The resulting dynamics allows us to investigate how the rotating Einstein--Gauss--Bonnet (EGB) black hole modifies the regular and chaotic regions of the particle's phase space. We employ Poincaré sections, the largest Lyapunov exponent (LLE), and the smaller alignment index (SALI) to characterise the orbital dynamics. We also examine the relationship between the maximum Lyapunov exponent and the surface gravity of the black hole, particularly in connection with the proposed near-horizon chaos bound.
\begin{figure*}
\centering 
\includegraphics[width=\hsize]{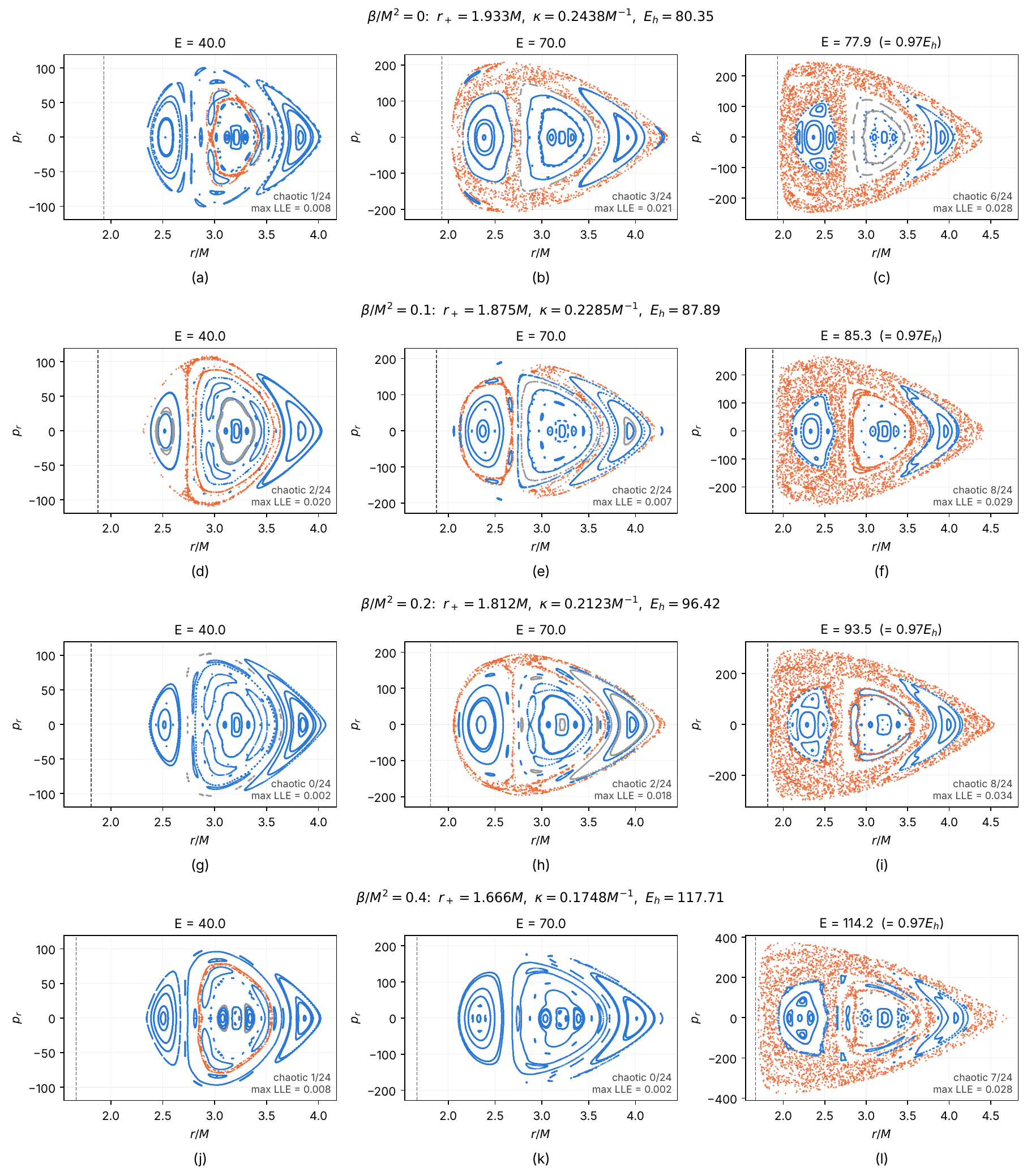}
\caption{Poincar\'e sections ($\theta=\pi/2$, $\dot\theta>0$) of the trapped charged
particle for $\beta/M^{2}=0$ (a) - (c), $0.1$ (d) - (f), $0.2$ (g) - (i) and $0.4$
(j) - (l), with $a=0.3M$ and $Q=0.2M$; $\beta=0$ is the Kerr--Newman case. The columns correspond to $E=40$, $E=70$ and $E=0.97\,E_h$, where $E_h$ is the trap energy at the outer horizon. Particle and trap parameters are $q=0.5$, $L=0$, $K_r=100$,
$K_\theta=25$ and $r_c=3.2M$. Each panel contains 24 orbits started at
$\theta=\pi/2$ with $p_r=0$ and $r$ distributed uniformly over the allowed interval,
coloured by their SALI at $t=5000M$: blue, regular (SALI $>10^{-4}$); orange, chaotic
(SALI $<10^{-8}$); grey, undecided. The dashed vertical line marks $r_+$. The values of
$r_+$, $\kappa$ and $E_h$ are given above each row; each panel gives the number of
chaotic orbits and the largest Lyapunov exponent among them (in units of $M^{-1}$).}\label{trappoincbeta}
\end{figure*}

\begin{figure*}
\centering 
\includegraphics[width=\hsize]{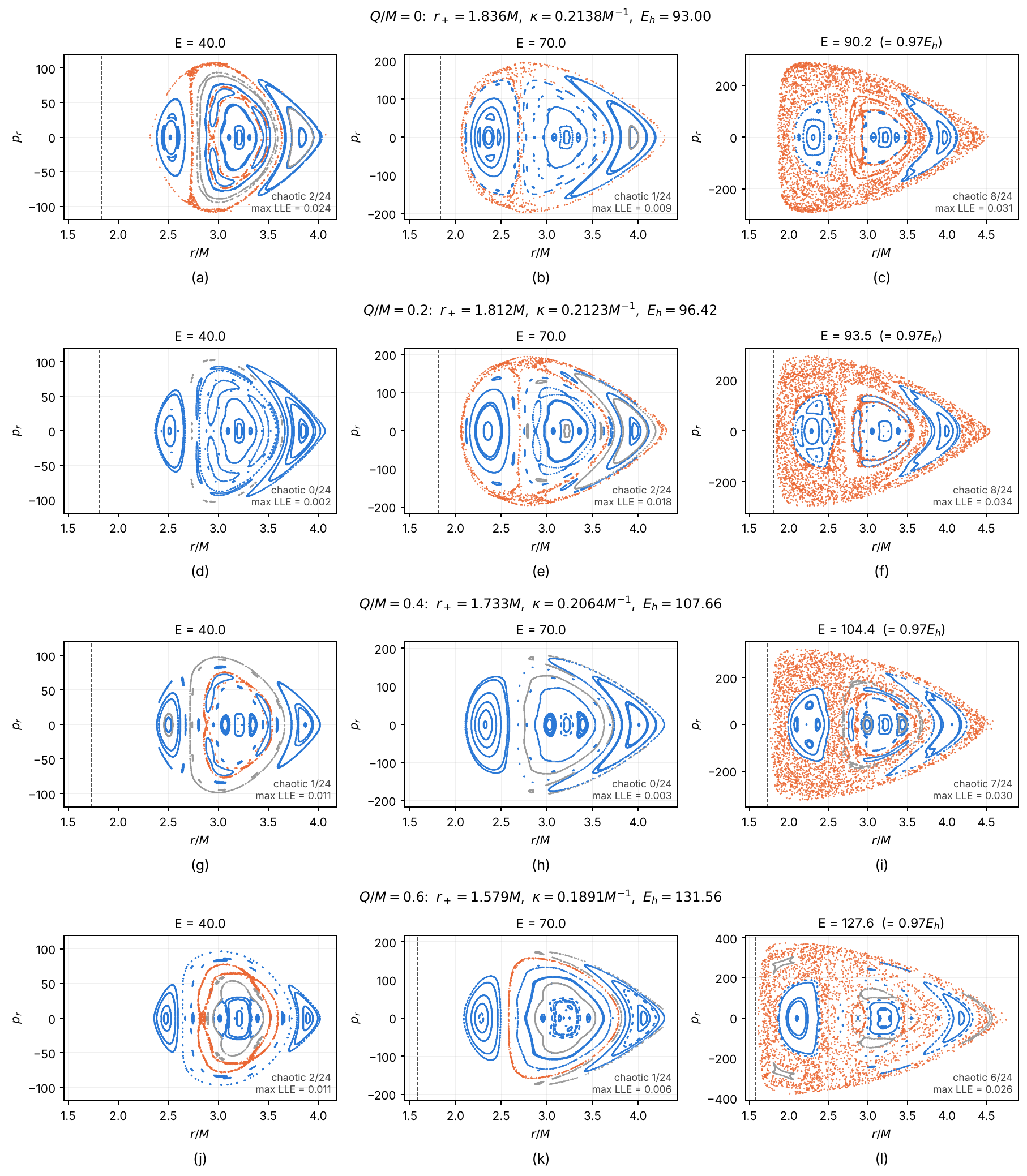}
\caption{As Fig.~\ref{trappoincbeta}, for $Q/M=0$ (a) - (c), $0.2$
(d) - (f), $0.4$ (g) - (i) and $0.6$ (j) - (l), with $a=0.3M$ and $\beta=0.2M^{2}$.
}\label{trappoincQ}
\end{figure*}

\begin{figure*}
\centering 
\includegraphics[width=\hsize]{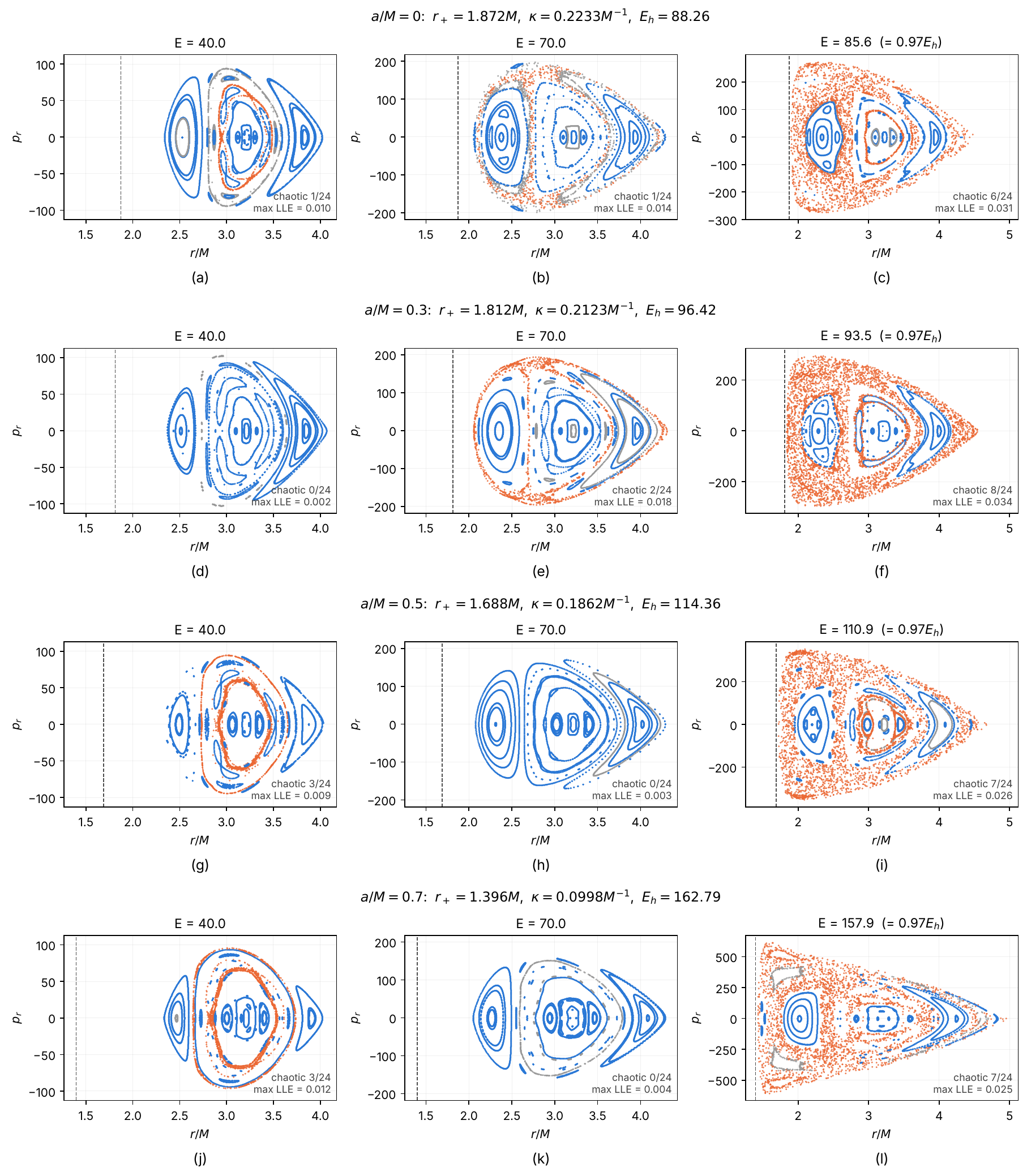}
\caption{As Fig.~\ref{trappoincbeta}, for $a/M=0$ (a) - (c), $0.3$
(d) - (f), $0.5$ (g) - (i) and $0.7$ (j) - (l), with $Q=0.2M$ and $\beta=0.2M^{2}$.
}\label{trappoinca}
\end{figure*}
\subsection{Harmonic trap and loss of separability}
We formulate the particle dynamics using the coordinate time $t$ as the evolution parameter. This choice is particularly useful because the Lyapunov exponent and the surface gravity $\kappa$ can then be compared in the same time parametrisation. Starting from the mass-shell condition and solving for the future-directed energy branch, $E=-\pi_t$, we obtain the energy Hamiltonian. The harmonic potential is subsequently incorporated into this Hamiltonian through the replacement of the particle energy by $E-V(r,\theta)$. The resulting Hamiltonian is given by,
\begin{equation}
\begin{split}
H={}&\omega L+qA_{t}'+V(r,\theta)\\
&+\alpha\sqrt{1+\frac{\left(L+qA_{\phi}\right)^{2}}{g_{\phi\phi}}
+\frac{\Delta\,p_{r}^{2}}{\rho^{2}}+\frac{p_{\theta}^{2}}{\rho^{2}}},
\end{split}
\label{eq:H_trap}
\end{equation}
 where the harmonic potential is defined as 
\begin{equation}
V(r,\theta)=\frac{K_{r}}{2}\left(r-r_{c}\right)^{2}
+\frac{K_{\theta}}{2}\,r_{c}^{2}\left(\theta-\frac{\pi}{2}\right)^{2},
\label{eq:trap}
\end{equation}
\begin{equation}
\begin{gathered}
\alpha^{2}=\frac{\rho^{2}\Delta}{\Sigma},\qquad
\omega=\frac{a\left(r^{2}+a^{2}-\Delta\right)}{\Sigma},\\
g_{\phi\phi}=\frac{\Sigma\sin^{2}\theta}{\rho^{2}},\qquad
A_{\phi}=-\frac{aQr\sin^{2}\theta}{\rho^{2}},\\
A_{t}'=A_{t}+\omega A_{\phi}=\frac{Qr\left(r^{2}+a^{2}\right)}{\Sigma},\\
\Sigma=\left(r^{2}+a^{2}\right)^{2}-\Delta a^{2}\sin^{2}\theta .
\end{gathered}
\label{eq:trap_defs}
\end{equation}

The motion follows from
\begin{equation}
\begin{gathered}
\frac{dr}{dt}=\frac{\partial H}{\partial p_{r}},\qquad
\frac{d\theta}{dt}=\frac{\partial H}{\partial p_{\theta}},\\
\frac{dp_{r}}{dt}=-\frac{\partial H}{\partial r},\qquad
\frac{dp_{\theta}}{dt}=-\frac{\partial H}{\partial \theta},
\end{gathered}
\label{eq:trap_eom}
\end{equation}
with $H=E$ conserved. Because the trap enters only through $-\pi_t=E-V(r,\theta)$, the
Hamilton--Jacobi equation is that of Sec.~\ref{S3} with
$E\to E-V(r,\theta)$:
\begin{equation}
\begin{split}
\Delta\left(\partial_{r}S\right)^{2}+\left(\partial_{\theta}S\right)^{2}
&-\frac{\left[(E-V)\left(r^{2}+a^{2}\right)-aL-qQr\right]^{2}}{\Delta}\\
&+\frac{\left[L-a(E-V)\sin^{2}\theta\right]^{2}}{\sin^{2}\theta}+\rho^{2}=0 .
\end{split}
\label{eq:HJ_trap}
\end{equation}
The quantities appearing in the Hamiltonian are expressed in terms of the metric and electromagnetic potential functions through Eqs.~(\ref{e6}-\ref{e7}). Here, $\omega$ represents the angular velocity associated with frame dragging, while $\alpha$ characterises the lapse function of the corresponding zero-angular-momentum observer. The azimuthal angular momentum $L=\pi_\phi$ remains conserved because the external potential is independent of the azimuthal coordinate. In particular, for $L=0$, the Hamiltonian describes the motion in the frame-dragging frame of a zero-angular-momentum observer, with the corresponding lapse function satisfying $F=\Delta\rho^2/\Sigma$.

The equations of motion are obtained from the canonical Hamilton equations, as presented in Eq.~(\ref{eq:H_trap}), with the Hamiltonian $H=E$ remaining conserved along each trajectory. The introduction of the harmonic potential modifies the energy dependence of the Hamilton--Jacobi equation through the replacement $E\rightarrow E-V(r,\theta)$, leading to Eq.~(\ref{eq:trap}). Since the potential depends simultaneously on the radial and polar coordinates, the radial and angular contributions can no longer be separated. Even a purely radial trapping potential generally introduces coupling between the two coordinates when $a\neq0$. Consequently, the Carter constant is no longer conserved, and the resulting dynamics can exhibit non-integrable behaviour.

To examine the resulting dynamics, we adopt the trapping parameters $K_r=100$, $K_\theta=25$, and $r_c=3.2M$, following Ref.~\cite{Dalui:2018qqv}. The angular displacement is defined as $y=r_c(\theta-\pi/2)$. Throughout this analysis, we consider a charged particle with specific charge $q=0.5$ and vanishing azimuthal angular momentum, $L=0$.

Before investigating the orbital dynamics, we verify the consistency of the Hamiltonian formulation. The numerical Hamiltonian reproduces the mass-shell condition with an accuracy of $4\times10^{-14}$. In the absence of the trapping potential, the Carter constant remains conserved to approximately $10^{-15}$ along the coordinate-time evolution. When a weak harmonic potential with $K_r=K_\theta=2\times10^{-3}$ is introduced, the Carter constant exhibits a variation of approximately $2.5\times10^{-3}$. These checks confirm that the trapping potential breaks the separability of the original particle dynamics.

An important quantity in the present analysis is the particle energy evaluated at the event horizon. We define $E_h$ as the limiting value of the Hamiltonian at $r=r_+$, on the equatorial plane and with vanishing canonical momenta. The condition $E<E_h$ provides the criterion for confined motion in the present trapping model. Furthermore, $E_h$ increases as the outer horizon radius decreases, allowing the particle to access higher energies before reaching the horizon.
\begin{equation}
\begin{split}
E_{h}&=\lim_{r\to r_{+}}H\!\left(r,\tfrac{\pi}{2},0,0\right)\\
&=-\frac{qQ\,r_{+}}{r_{+}^{2}+a^{2}}+\frac{K_{r}}{2}\left(r_{c}-r_{+}\right)^{2}
\quad (L=0).
\end{split}
\label{eq:Eh}
\end{equation}
\begin{figure*}
\centering 
\includegraphics[width=\hsize]{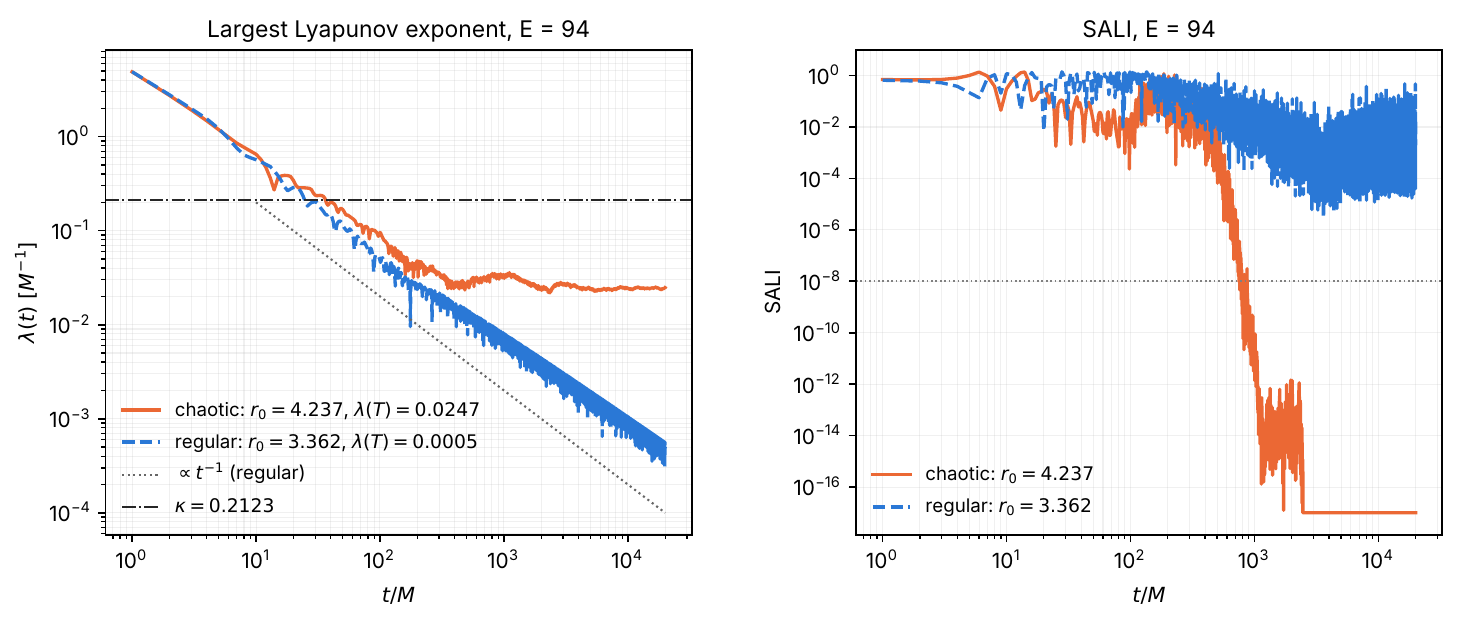}
\caption{Finite-time largest Lyapunov exponent $\lambda(t)$ (left) and SALI (right) at
$E=94$ for a chaotic orbit ($r_0=3.690M$, solid) and a regular orbit ($r_0=3.252M$,
dashed); black hole.
Dash-dotted line: $\kappa=0.2123M^{-1}$; dotted lines: $t^{-1}$ reference (left) and
SALI $=10^{-8}$ (right).}\label{trapsalitime}
\end{figure*}

\begin{figure*}
\centering 
\includegraphics[width=\hsize]{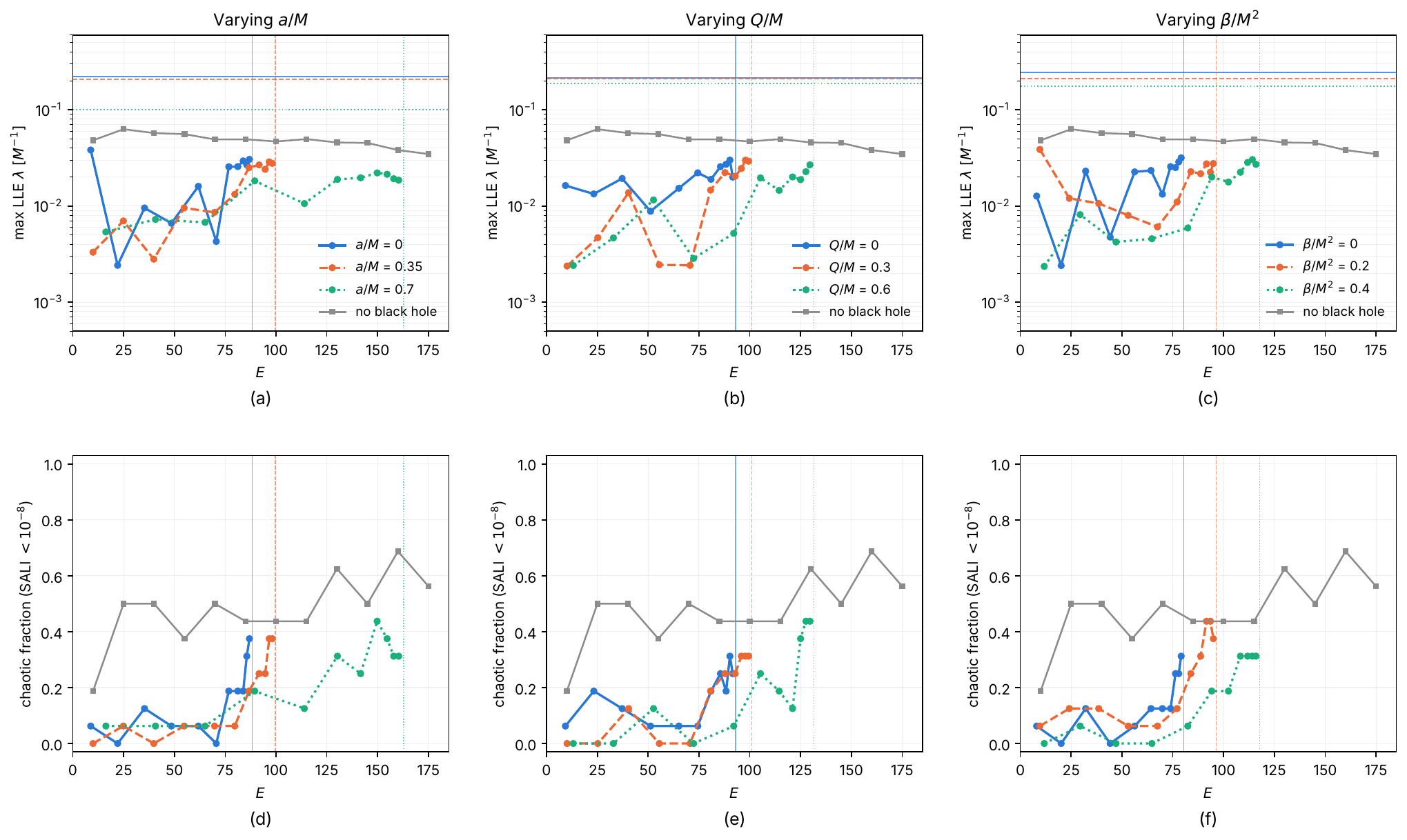}
\caption{Largest Lyapunov exponent maximised over 16 orbits, (a) - (c), and fraction of
chaotic orbits (SALI $<10^{-8}$ at $t=5000M$), (d) - (f), versus $E$ for varying $a$,
$Q$ and $\beta$, with the other two fixed at $a=0.3M$, $Q=0.2M$, $\beta=0.2M^{2}$.
Horizontal and vertical lines mark $\kappa$ and $E_h$ in the matching colour; grey
squares: no black hole.}\label{trapsaliscan}
\end{figure*}
\subsection{Poincar\'e sections}
The phase-space structures associated with variations in the black hole parameters are presented in Figs.~\ref{trappoincbeta}--\ref{trappoinca}. The three columns correspond to $E=40$, $E=70$, and $E=0.97E_h$, respectively. Each panel contains 24 trajectories, allowing us to compare the regular and chaotic regions at different energy scales.

We first consider the effect of the Gauss--Bonnet coupling. At $E=70$, the number of chaotic trajectories decreases from three for $\beta=0$ to two for $\beta=0.1M^2$ and $\beta=0.2M^2$, and none for $\beta=0.4M^2$. The corresponding maximum LLE decreases from $0.021M^{-1}$ to $0.002M^{-1}$, although not monotonically ($0.007M^{-1}$ for $\beta=0.1M^2$ and $0.018M^{-1}$ for $\beta=0.2M^2$). This behaviour indicates a suppression of chaotic motion at fixed energy as the coupling increases. However, when the energy is expressed as a fraction of the horizon energy, the situation changes. At $E=0.97E_h$, between six and eight trajectories are chaotic across the considered coupling values, with maximum LLEs ranging from $0.028M^{-1}$ to $0.034M^{-1}$. The chaotic sea therefore remains prominent close to the horizon for all the investigated values of $\beta$.

The variation of the electric charge produces a weaker but similar trend. At $E=70$, one and two trajectories are chaotic for $Q=0$ and $Q=0.2M$, respectively, whereas none is found for $Q=0.4M$ and one for $Q=0.6M$. The maximum LLE correspondingly decreases from $0.009$--$0.018M^{-1}$ for the two smaller charges to $0.003$--$0.006M^{-1}$ for the two larger ones. At $E=0.97E_h$, however, six to eight trajectories are classified as chaotic, with maximum LLE values between $0.026M^{-1}$ and $0.034M^{-1}$. Thus, although increasing the charge weakens the chaotic contribution at a fixed energy, the near-horizon chaotic region remains comparable when the energy is normalised by $E_h$.

The influence of rotation follows the same general pattern. At $E=70$, one trajectory is chaotic for $a=0$ and two for $a=0.3M$, whereas none is found for $a=0.5M$ and $a=0.7M$, and the maximum LLE drops from $0.014$--$0.018M^{-1}$ to $0.003$--$0.004M^{-1}$. At $E=0.97E_h$, six to eight trajectories are chaotic across the investigated spin values, with maximum LLEs between $0.025M^{-1}$ and $0.034M^{-1}$. We also observe that the radial momenta can reach approximately $|p_r|\approx600$ for $a=0.7M$ near the smaller horizon, indicating a substantially different momentum scale in this regime.

At the lower energy $E=40$, the Poincar\'e sections predominantly exhibit regular island structures accompanied by thin chaotic layers. Their distribution does not follow a systematic trend with increasing $\beta$, $Q$, or $a$. For example, three chaotic trajectories are found for $a=0.5M$ and $a=0.7M$ and two for $Q=0.6M$, while none is obtained for $a=0.3M$. This indicates that the low-energy phase-space structure is sensitive to the detailed orbital resonances rather than following a simple monotonic dependence on the black hole parameters.

This detailed analysis shows that the apparent suppression of chaos with increasing $a$, $Q$, and $\beta$ at fixed energy is closely related to the corresponding increase in the horizon energy. Once the energy is expressed relative to $E_h$, the chaotic fractions and maximum Lyapunov exponents become comparable across the different configurations. The near-horizon region therefore provides a common regime in which chaotic dynamics becomes prominent, largely irrespective of the individual black hole parameters considered here.

\subsection{Chaos indicators: LLE and SALI}
\label{sec:indicators}

The Poincar\'e sections identify chaotic regions qualitatively. To quantify the
sensitivity to initial conditions we use two indicators computed from the tangent
dynamics: the largest Lyapunov exponent (LLE)~\cite{benettin1980lyapunov} and the
smaller alignment index (SALI)~\cite{Skokos:2004mt}. Writing the phase-space point as
$\mathbf{x}=(r,\theta,p_r,p_\theta)^{\mathsf T}$, the equations of
motion~(\ref{eq:trap_eom}) take the form
\begin{equation}
\frac{d\mathbf{x}}{dt}=\mathbf{F}(\mathbf{x})=\mathsf{\Omega}\,\nabla_{\mathbf{x}}H,
\qquad
\mathsf{\Omega}=\begin{pmatrix}\mathsf{0}&\mathsf{I}_{2}\\-\mathsf{I}_{2}&\mathsf{0}\end{pmatrix},
\label{eq:flow}
\end{equation}
where $H$ is the Hamiltonian~(\ref{eq:H_trap}) and the coordinate time $t$ is the
evolution parameter. A small deviation $\mathbf{w}=\delta\mathbf{x}$ from the orbit
evolves according to the variational equations
\begin{equation}
\frac{d\mathbf{w}}{dt}=\mathsf{J}\big(\mathbf{x}(t)\big)\,\mathbf{w},\qquad
\mathsf{J}=\frac{\partial\mathbf{F}}{\partial\mathbf{x}}
=\mathsf{\Omega}\,\frac{\partial^{2}H}{\partial\mathbf{x}\,\partial\mathbf{x}^{\mathsf T}},
\label{eq:variational}
\end{equation}
which are integrated together with Eq.~(\ref{eq:flow}).

\emph{Largest Lyapunov exponent.} The LLE is the mean exponential rate at which
neighbouring orbits separate,
\begin{equation}
\lambda=\lim_{t\to\infty}\frac{1}{t}\,
\ln\frac{\left\|\mathbf{w}(t)\right\|}{\left\|\mathbf{w}(0)\right\|},
\label{eq:lle_def}
\end{equation}
which is positive for a chaotic orbit and zero for a regular one. Because
$\|\mathbf{w}\|$ grows exponentially along a chaotic orbit, we evaluate
Eq.~(\ref{eq:lle_def}) with the renormalisation method of Benettin
\textit{et al.}~\cite{benettin1980lyapunov}: the deviation vector $\mathbf{w}_1$ is
rescaled to unit length at the times $t_k=k\,\Delta t$, and the finite-time estimate is
\begin{equation}
\lambda(t_{n})=\frac{1}{t_{n}}\sum_{k=1}^{n}\ln\alpha_{k},\qquad
\alpha_{k}=\left\|\mathbf{w}_{1}(t_{k})\right\|,
\label{eq:lle}
\end{equation}
where $\alpha_k$ is the norm reached at $t_k$, before rescaling, by a vector of unit
length at $t_{k-1}$. For a chaotic orbit $\lambda(t_n)$ settles on a positive plateau,
whereas for a regular orbit, along which deviations grow only linearly in time, it
decays approximately as $\ln t/t$.

\emph{Smaller alignment index.} The SALI follows two deviation vectors
$\mathbf{w}_1$ and $\mathbf{w}_2$, initially orthonormal, and measures whether they
become aligned,
\begin{equation}
\mathrm{SALI}(t)=\min\left\{\left\|\hat{\mathbf{w}}_{1}(t)+\hat{\mathbf{w}}_{2}(t)\right\|,\,
\left\|\hat{\mathbf{w}}_{1}(t)-\hat{\mathbf{w}}_{2}(t)\right\|\right\},\qquad
\hat{\mathbf{w}}_{i}=\frac{\mathbf{w}_{i}}{\left\|\mathbf{w}_{i}\right\|},
\label{eq:sali}
\end{equation}
so that $0\le\mathrm{SALI}\le\sqrt{2}$. Along a chaotic orbit both vectors turn towards
the most unstable direction and become parallel or antiparallel, so the SALI falls
exponentially, as $\mathrm{SALI}\propto\exp[-(\lambda_1-\lambda_2)t]$ with
$\lambda_1\ge\lambda_2$ the two largest Lyapunov exponents~\cite{Skokos:2004mt}. For
the present system with two degrees of freedom $\lambda_2=0$, so that
$\mathrm{SALI}\propto e^{-\lambda t}$. Along a regular orbit the two vectors become
tangent to the invariant torus without aligning, and the SALI fluctuates around a
non-zero value. The SALI therefore separates the two types of orbit well before
$\lambda(t)$ has converged. We classify an orbit as chaotic if
$\mathrm{SALI}<10^{-8}$ and as regular if $\mathrm{SALI}>10^{-4}$ at the end of the
integration, and as undecided otherwise.

\emph{Numerical procedure.} Equations~(\ref{eq:flow}) and (\ref{eq:variational}) are
integrated together with an adaptive Dormand--Prince 5(4) scheme, with relative and
absolute tolerances of $10^{-10}$ and $10^{-11}$, starting from the orthonormal pair
$\mathbf{w}_1(0)=\tfrac12(1,1,1,1)^{\mathsf T}$ and
$\mathbf{w}_2(0)=\tfrac12(1,-1,1,-1)^{\mathsf T}$. Both vectors are rescaled to unit
length every $\Delta t=1M$, where $\lambda(t_n)$ and $\mathrm{SALI}(t_n)$ are recorded.
All norms are Euclidean in the coordinates $(r,\theta,p_r,p_\theta)$. Orbits are
classified at $T=5000M$, and the representative orbits discussed below are followed to
$T=2\times10^{4}M$. Since $\lambda$ is measured with respect to the coordinate time
$t$, the same time with respect to which the surface gravity is defined, $\lambda$ and
$\kappa$ can be compared directly. The initial conditions lie on the section with
$\theta_0=\pi/2$ and $p_{r0}=0$, at equally spaced radii $r_0$ inside the
energetically allowed interval, and $p_{\theta0}\ge0$ is fixed by $H=E$:
\begin{equation}
\begin{split}
p_{\theta0}^{2}=\rho^{2}\Bigg[&\frac{\left(E-\omega L-qA_{t}'-V\right)^{2}}{\alpha^{2}}\\
&-1-\frac{\left(L+qA_{\phi}\right)^{2}}{g_{\phi\phi}}\Bigg]_{\theta=\pi/2,\ p_r=0}.
\end{split}
\label{eq:ic}
\end{equation}

\subsection{Time evolution of the LLE and SALI}
\label{sec:lle_sali_time}

We first examine the time evolution of the two chaos indicators for representative regular and chaotic trajectories. Fig.~\ref{trapsalitime} displays their evolution at $E=94$ for the black hole with $a=0.3M$, $Q=0.2M$ and $\beta=0.2M^2$. The chaotic trajectory, with initial radial coordinate $r_0=4.237M$, exhibits a rapid increase in the finite-time LLE followed by an approximately constant plateau. At $T=2\times10^4M$, the exponent reaches
$\lambda=0.0247M^{-1}=0.117\kappa$.
The corresponding SALI decreases exponentially and falls below $10^{-15}$ after approximately $1100M$. This rapid alignment of the deviation vectors provides a clear indication of exponential sensitivity to initial conditions.

The regular trajectory, with $r_0=3.362M$, exhibits a different behaviour. Its finite-time LLE decreases approximately as $\ln t/t$, as expected for a regular orbit, and reaches $5.3\times10^{-4}M^{-1}$ at the end of the integration. Unlike the chaotic case, the SALI does not show a sustained exponential decay. Instead, it fluctuates between approximately $3\times10^{-6}$ and $0.5$ for $t>1000M$. The contrasting behaviour of the two indicators consistently distinguishes the regular and chaotic trajectories. In particular, the SALI identifies the chaotic character of the orbit at relatively early times, around $10^3M$, whereas the LLE requires a substantially longer integration interval to approach its asymptotic value.

\subsection{Dependence on energy and black-hole parameters}
\label{sec:lle_sali_scan}

We further investigate the influence of the black hole parameters on the onset and strength of chaotic motion. Fig.~\ref{trapsaliscan} presents the energy-dependent behaviour of the LLE and SALI for different values of the spin parameter $a$, electric charge $Q$, and Gauss--Bonnet coupling $\beta$. For each configuration, we determine the surface gravity, horizon energy, maximum LLE, and fraction of chaotic trajectories.

For the spin variation, the surface gravity decreases from $0.2233M^{-1}$ at $a=0$ to $0.0998M^{-1}$ at $a=0.7M$. Over the same range, the horizon energy increases from $88.26$ to $162.79$. The maximum LLE in the near-horizon range, defined here by $E\geq0.92E_h$, decreases from $0.031M^{-1}$ to $0.022M^{-1}$, while the ratio $\lambda/\kappa$ increases from $0.14$ to $0.22$. The chaotic fraction in this range increases from $0.27$ to $0.36$. At lower energies, $E\leq0.4E_h$, the corresponding fraction remains between $0.02$ and $0.06$.

A similar behaviour is observed when the electric charge is varied. Increasing $Q$ from zero to $0.6M$ reduces the surface gravity from $0.2138M^{-1}$ to $0.1891M^{-1}$ and increases $E_h$ from $93.00$ to $131.56$. The maximum LLE remains within the range $0.027$--$0.030M^{-1}$, while $\lambda/\kappa\approx0.14$ for all three charges. The chaotic fraction near the horizon increases from $0.25$ to $0.34$, whereas at lower energies it remains between $0.04$ and $0.12$.

The influence of the Gauss--Bonnet coupling is also evident. For $\beta=0$, the surface gravity and horizon energy are $0.2438M^{-1}$ and $80.35$, respectively. Increasing the coupling to $\beta=0.4M^2$ changes these values to $0.1748M^{-1}$ and $117.71$. The maximum LLE near the horizon varies between $0.028M^{-1}$ and $0.032M^{-1}$, with the corresponding ratio $\lambda/\kappa$ remaining between $0.13$ and $0.17$. For $\beta=0.2M^2$, the largest value obtained over the complete scan is $0.039M^{-1}$, from a thin chaotic layer at $E\approx10$, while the maximum LLE restricted to $E\geq0.92E_h$ is $0.028M^{-1}$; a similar low-energy layer gives $0.038M^{-1}$ for $a=0$. The chaotic fraction near the horizon increases from $0.23$ at $\beta=0$ to $0.39$ at $\beta=0.2M^2$, before decreasing to $0.31$ at $\beta=0.4M^2$.

Across all the investigated configurations, the chaotic fraction remains relatively small at low energies, ranging from $0.02$ to $0.12$. As the energy approaches the horizon value, it increases substantially and reaches approximately $0.23$--$0.39$. The largest Lyapunov exponents in this near-horizon regime are of the order of $0.02$--$0.03M^{-1}$. These results establish a clear association between the accessibility of the near-horizon region and the enhancement of chaotic motion.

For comparison, the control system without a black hole exhibits larger LLE values, ranging from $0.035M^{-1}$ to $0.063M^{-1}$, together with a chaotic fraction of approximately $0.48$. Unlike the black hole configurations, this system does not exhibit a clear systematic energy dependence in its chaotic fraction. This comparison further demonstrates that the black hole primarily modifies the distribution and strength of the chaos already supported by the trapping potential.

An additional observation concerns the relationship between the LLE and the surface gravity. For all the considered parameter configurations, the ratio $\lambda/\kappa$ remains within the range $0.13$--$0.22$. Thus, the calculated exponents satisfy the proposed near-horizon bound $\lambda\leq\kappa$~\cite{Hashimoto2017Universality,Maldacena:2015waa} with a substantial margin. The largest ratio occurs for $a=0.7M$, where the surface gravity attains its smallest value among the investigated spin configurations. The dependence on $a$, $Q$, and $\beta$ is therefore associated with changes in the horizon radius, horizon energy, and surface gravity, which collectively influence the accessibility of the near-horizon region.


\section{Local and trajectory exponents}\label{s_disc}

The two parts of this work use the Lyapunov exponent in different senses. In
Sec.~\ref{s3} the motion is integrable, and $\lambda$ is the local exponent of an
isolated unstable equilibrium: it measures how fast nearby orbits leave that
equilibrium, not chaos. In Sec.~\ref{S4} the trap breaks integrability, and $\lambda$
is the trajectory LLE of Eq.~(\ref{eq:lle}), a time average of the stretching along a
chaotic orbit. Both are compared with the same surface gravity $\kappa$, so it is
natural to ask how they are related. To answer this we compute, for the nine black holes
of Fig.~\ref{trapsaliscan} and $L=0$, the local exponents of both systems and compare
them with the trajectory LLE of the trap.

On the equator, with $p_r=p_\theta=0$, the energy Hamiltonian~(\ref{eq:H_trap}) reduces
to the radial potential
\begin{equation}
U(r)=\omega L+qA_t'+\sqrt{F\left(1+X\right)}+V(r),\qquad
X=\frac{\left(L+qA_\phi\right)^{2}}{D},
\label{eq:U_local}
\end{equation}
with $F=\Delta r^{2}/\Sigma$, $N=\Delta/r^{2}$ and $D=\Sigma/r^{2}$ evaluated at
$\theta=\pi/2$, and $V(r)=\tfrac12K_r(r-r_c)^2$ with the trap or $V=0$ without it. An
equilibrium $r_e$ satisfies $U'(r_e)=0$, and its radial exponent is
\begin{equation}
\lambda_{\rm loc}^{2}=-U''(r_e)\,\frac{F N}{\sqrt{F\left(1+X\right)}} ,
\label{eq:lambda_local}
\end{equation}
which for $V=0$ is the exponent of Sec.~\ref{s3}. With the trap, the polar direction
is stable ($K_\theta>0$), so $r_e$ is a saddle-centre and Eq.~(\ref{eq:lambda_local})
gives its only unstable rate. The results are collected in Table~\ref{tab:local_global}.

\begin{table*}[ht]
\centering
\caption{Local and trajectory Lyapunov exponents for the nine black holes of
Fig.~\ref{trapsaliscan} ($L=0$, $M=1$). Each row varies one parameter of the reference
black hole $a=0.3M$, $Q=0.2M$, $\beta=0.2M^2$. $\lambda_{\rm loc}$ is the radial
exponent, Eq.~(\ref{eq:lambda_local}), at the first unstable equilibrium outside the
horizon: for $q=10$ without the trap (method of Sec.~\ref{s3}) and for $q=0.5$ with the
trap, where the equilibrium lies at a distance $r_e-r_+$ from the horizon. A $q=0.5$
particle has no equilibrium without the trap. $\lambda_{\rm max}$ is the largest
trajectory LLE of the trap for $E\ge0.92E_h$. A dash means that no equilibrium
exists.}
\label{tab:local_global}
\begin{tabular}{|c|c|c|c|c|c|c|}
\hline
Black hole & $r_+$ & $\kappa$ & $\lambda_{\rm loc}/\kappa$ ($q=10$) & $r_e-r_+$ (trap) & $\lambda_{\rm loc}/\kappa$ (trap) & $\lambda_{\rm max}/\kappa$ (trap) \\
\hline
$a=0$ & 1.8718 & 0.2233 & 0.6243 & $6.3\times10^{-6}$ & 0.999995 & 0.137 \\
$a=0.35$ & 1.7886 & 0.2078 & 0.6583 & $5.0\times10^{-6}$ & 0.999996 & 0.139 \\
$a=0.7$ & 1.3959 & 0.0998 & 1.1050 & $1.2\times10^{-6}$ & 0.999999 & 0.222 \\
$Q=0$ & 1.8362 & 0.2138 & -- & $5.6\times10^{-6}$ & 0.999996 & 0.140 \\
$Q=0.3$ & 1.7800 & 0.2101 & 0.8626 & $5.1\times10^{-6}$ & 0.999996 & 0.142 \\
$Q=0.6$ & 1.5791 & 0.1891 & 0.9889 & $3.5\times10^{-6}$ & 0.999998 & 0.142 \\
$\beta=0$ & 1.9327 & 0.2438 & 0.5851 & $7.4\times10^{-6}$ & 0.999994 & 0.130 \\
$\beta=0.2$ & 1.8117 & 0.2123 & 0.6482 & $5.4\times10^{-6}$ & 0.999996 & 0.130 \\
$\beta=0.4$ & 1.6661 & 0.1748 & 0.7256 & $3.6\times10^{-6}$ & 0.999998 & 0.174 \\
\hline
\end{tabular}
\end{table*}

\emph{Particle without the trap ($q=10$).} The fourth column links the table to
Sec.~\ref{s3}. For $Q\neq0$ the Coulomb repulsion balances gravity at an unstable
equilibrium, and $\lambda_{\rm loc}/\kappa$ lies between $0.59$ and $0.99$ for seven of
the eight charged black holes. It increases with $a$, $Q$ and $\beta$, in line with the
rise of $\lambda^2-\kappa^2$ found in Sec.~\ref{s3}, and for $a=0.7M$ it reaches
$1.105$, so the bound is violated already at $L=0$. For $Q=0$ there is no Coulomb force
and hence no equilibrium.

\emph{Trapped particle ($q=0.5$).} A particle with $q=0.5$ has no equilibrium without
the trap, because its Coulomb repulsion is too weak to balance gravity. With the trap,
the outward trap force $-V'(r)=K_r(r_c-r)$ balances gravity at a point extremely close
to the horizon, $r_e-r_+$ between $1.2\times10^{-6}M$ and $7.4\times10^{-6}M$, because
the force needed to hold a particle at rest diverges as $r\to r_+$. At this point
$\lambda_{\rm loc}/\kappa$ lies between $0.999994$ and $0.999999$, and $1-\lambda_{\rm loc}/\kappa$
is of the same order as $(r_e-r_+)/M$. The local exponent therefore saturates the bound
as the equilibrium approaches the horizon, in agreement with the near-horizon
universality of Ref.~\cite{Hashimoto2017Universality}.

\emph{Trajectory exponent.} For the same trapped particle, the largest trajectory LLE
near $E_h$ is only $0.13\kappa$ - $0.22\kappa$, between about 4.5 and 7.7 times
smaller than the local exponent at the near-horizon saddle. It is largest for $a=0.7M$, where
$\kappa$ is smallest. We attribute this gap to the fact that the LLE averages the
stretching rate over the whole orbit, whereas the rate approaches $\kappa$ only within
a very thin layer next to the horizon, in which chaotic orbits spend a small fraction
of their time; we have not measured this fraction directly.

Taken together, the table gives the ordering
\begin{equation}
\lambda_{\rm max}<\lambda_{\rm loc}(q=10)<\lambda_{\rm loc}({\rm trap})\simeq\kappa
\label{eq:exponent_ordering}
\end{equation}
for every black hole except $a=0.7M$, where $\lambda_{\rm loc}(q=10)$ exceeds $\kappa$,
and $Q=0$, where the $q=10$ particle has no equilibrium.
The two parts of the paper are therefore complementary. Section~\ref{s3} shows where
the local exponent of an isolated equilibrium can exceed $\kappa$, which requires a
strongly charged particle in an integrable system, while Sec.~\ref{S4} shows that the
rate of genuinely chaotic motion stays well below $\kappa$. In the present setup,
values of $\lambda$ above $\kappa$ thus describe the local instability of equilibria
rather than the growth rate of chaos. Since the $q=10$ and trap columns refer to
different particles, the most direct comparison is between the last two columns, which
belong to the same system.

\subsection*{Assumptions and limitations}\label{s5}

The results above hold under the following assumptions, which we state so that they
are not over-read.
\begin{enumerate}
\item \emph{Effective background.} The rotating metric is obtained from the static 4D EGB solution by the Newman--Janis algorithm, which does not in general generate
solutions of modified gravity theories~\cite{Hansen:2013owa}, and we do not claim
that it solves the field equations of any four-dimensional EGB theory. The coupling
$\beta$ is an effective parameter of this background. The $D\rightarrow4$ limit of
Glavan and Lin is itself debated, and consistent regularisations exist; we do not discuss which of them, if any, the rotating metric belongs to.
\item \emph{Probe approximation.} The particle is a test particle. Back-reaction on the metric, the particle's own electromagnetic field and radiation reaction are neglected. The charge-to-mass ratio $q=10$ used in Sec.~\ref{s3} is large, so the probe description there should be regarded as a model of orbital stability, not as a
statement about a physical particle. The trap study of Sec.~\ref{S4} uses $q=0.5$ and $L=0$ only.
\item \emph{Energy conditions and stability.} We restrict to parameters for which the horizon exists and $\Delta(r)$ is real, that is, $1+4\beta\left(2M/r^3-Q^2/r^4\right)>0$ outside the horizon. We have not checked the energy conditions of the effective matter content of the Newman--Janis metric, nor the dynamical stability of the black hole itself.
\item \emph{Higher-order corrections.} Terms beyond the Gauss--Bonnet order, as well
as quantum and string-scale corrections, are not included. We therefore make no
statement about whether such corrections would restore $\lambda\le\kappa$.
\item \emph{Status of the bound.} For a classical probe the bound $\lambda\le\kappa$
is a conjecture inherited from the quantum chaos bound, and for $\kappa=0$ it is not
applicable. The local exponents of Sec.~\ref{s3} and the trajectory exponents of
Sec.~\ref{S4} are different quantities, and only the latter measure chaos.
\item \emph{Trap model.} The harmonic trap is a phenomenological external force, not a gravitational interaction. It is defined in Boyer--Lindquist coordinates, so it is not coordinate-invariant, and the dynamics depends on its functional form, on the constants $K_r$, $K_\theta$ and $r_c$, and on the factor $r_c^2$ that scales the
angular displacement.
\item \emph{Numerical estimates.} The LLEs are finite-time estimates, computed with
respect to the coordinate time $t$ and with the Euclidean norm in the phase-space
coordinates $(r,\theta,p_r,p_\theta)$, and their values should be interpreted within
these conventions. The chaotic fractions near $E_h$ are based on 64 orbits per case
(16 at each of four energies), for which a binomial estimate of the sampling
uncertainty is about $0.06$, so differences smaller than about $0.1$ between cases are not significant.
\end{enumerate}
\section{Conclusion}\label{S5}
We studied the motion of a charged test particle around the rotating charged black
hole of four-dimensional EGB gravity, with $M=1$, for varying charge $Q$, spin $a$,
GB coupling $\beta$ and particle angular momentum $L$. The work has two parts: the
local instability of equilibrium orbits and its comparison with the bound
$\lambda\le\kappa$, for a particle with charge-to-mass ratio $q=10$
(Sec.~\ref{s3}), and the chaotic dynamics of a particle with $q=0.5$ and $L=0$
confined near the horizon by a harmonic trap (Sec.~\ref{S4}). In both parts
$qQ>0$, so that the Coulomb force is repulsive. Our main conclusions are as follows.

\emph{Integrability.} For any metric function $\Delta(r)$ the Hamilton--Jacobi
equation of a charged particle separates, so the motion admits the Carter constant and
is integrable (Sec.~\ref{S3}). Poincar\'e sections of orbits near the ISCO consist
of smooth invariant curves for all values of $\beta$, $Q$ and $a$ considered
(Fig.~\ref{fig:poincare_isco}). Without an external perturbation the motion is
therefore not chaotic, and the Lyapunov exponent of an equilibrium orbit measures only
its local instability. These equilibria and exponents must be obtained from the exact
energy Hamiltonian, which retains the frame-dragging term $\omega L$; omitting this
term changes both for $L\neq0$.

\emph{Non-extremal black holes.} The equilibrium radius is smallest at $L=0$ for
$a=0$ and for slightly co-rotating orbits ($L\approx1$ - $2$) for $a\neq0$, and it
grows with $|L|$ away from this minimum. It decreases with increasing $Q$ and $\beta$,
and with increasing $a$ except for strongly counter-rotating orbits ($L\lesssim-2$),
for which it changes little (Tables~\ref{t1}, \ref{t2} and \ref{t3}). At fixed $L$,
increasing $Q$, $a$ or $\beta$ raises $\lambda^2-\kappa^2$. For $a=0$ it is negative
for all $L$, with a shallow maximum near $L\approx3$ - $5$ (Fig.~\ref{f1}). For
rotating black holes it is smallest near $L\approx0$ - $1$ and rises towards negative
$L$. Co-rotating orbits ($L\ge1$) satisfy $\lambda\le\kappa$ in every case examined,
whereas orbits with $L\le0$, mainly counter-rotating ones, violate it once $a$, $Q$ or
$\beta$ is large enough. For $Q=0.3$ and $\beta=0.3$ the bound holds for $a=0.3$ and
is violated for $L\lesssim-1.6$ at $a=0.4$ and for $L\lesssim-0.6$ at $a=0.5$, where
$\lambda^2-\kappa^2$ reaches $0.0051$ and $0.0136$ at $L=-3$ (Fig.~\ref{f2}). For
$a=Q=0.3$ the bound is violated only for $\beta=0.4$ and $L\lesssim-2$, with
$\lambda^2-\kappa^2=0.0033$ at $L=-3$ (Fig.~\ref{f3}). For $a=0.7$, $Q=0.2$ and
$\beta=0.2$ the violation already occurs at $L=0$, where $\lambda_{\rm loc}/\kappa=1.105$
(Table~\ref{tab:local_global}). The GB coupling therefore moves counter-rotating
orbits towards violation at fixed $Q$ and $a$, but does not make co-rotating orbits
violate the bound within the range examined. The event-horizon radius and the surface
gravity both decrease as $Q$, $a$ or $\beta$ increases.

\emph{Extremal black holes.} Here $\kappa=0$ and the bound cannot be tested, because
$\lambda^2>0$ for any unstable orbit; we use $\lambda^2$ only as a diagnostic of
instability. Counter-rotating particles have unstable equilibria, with $\lambda^2$ growing
as $L$ becomes more negative, up to $0.045$ for $\beta=0.1$ (Table~\ref{t5},
Fig.~\ref{f5}) and $0.034$ for $Q=0.6$ (Table~\ref{t7}, Fig.~\ref{f7}). Co-rotating
particles have no equilibrium outside the horizon. For $Q=0.6$, increasing $\beta$
(which lowers the extremal spin) lowers $\lambda^2$ and removes the equilibria first at
the smallest $|L|$.

\emph{Trapped particle.} The harmonic trap destroys the Carter constant, and the
trapped motion becomes chaotic as the particle energy approaches its value at the
horizon, $E_h$. The parameters $a$, $Q$ and $\beta$ act
mainly through $r_+$, $E_h$ and $\kappa$: at equal $E/E_h$ the degree of chaos is
similar, whereas at fixed $E$ the largest values give the most regular sections,
although not monotonically, because $E_h$ increases (Figs.~\ref{trappoincbeta}, \ref{trappoincQ} and \ref{trappoinca}). The LLE and the SALI classify the
orbits consistently, and the SALI separates them by $t\sim10^3M$, well before the LLE
has converged (Fig.~\ref{trapsalitime}). Over the parameter scan the chaotic
fraction rises from between $0.02$ and $0.12$ at $E\le0.4E_h$ to between $0.23$ and
$0.39$ for $E\ge0.92E_h$ (Fig.~\ref{trapsaliscan}).  The
largest LLE for $E\ge0.92E_h$ lies between $0.022M^{-1}$ and $0.032M^{-1}$, that is,
between $0.13\kappa$ and $0.22\kappa$, so the trajectory exponents satisfy
$\lambda\le\kappa$ by a wide margin. In the control calculation without the black hole
($M=0$, $a=Q=0$) the trap is itself chaotic at these energies, with a chaotic fraction
of about $0.48$ and largest LLE between $0.035M^{-1}$ and $0.063M^{-1}$, both larger
than with the black hole. The black hole therefore does not create the chaos of this
system; it changes its strength and concentrates it near the horizon.

\emph{Local and trajectory exponents.} With the trap, the first unstable equilibrium
lies only $(1$ - $7)\times10^{-6}M$ outside the horizon, and its local exponent
satisfies $\lambda_{\rm loc}/\kappa=0.999994$ - $0.999999$, in agreement with the
near-horizon universality of Ref.~\cite{Hashimoto2017Universality}. The trajectory
LLE of the same system is nevertheless only $0.13\kappa$ - $0.22\kappa$, which we
attribute to the short time chaotic orbits spend near this equilibrium. Without the
trap, a $q=10$ particle with $L=0$ has $\lambda_{\rm loc}/\kappa$ between $0.59$ and
$0.99$ for seven of the eight charged black holes of Table~\ref{tab:local_global}, and
$1.105$ for $a=0.7$. In the present setup, values of $\lambda$ above $\kappa$ are therefore a
property of isolated equilibria of strongly charged particles, whereas genuinely
chaotic trajectories remain well within the bound.

This study separates local instability from chaos. Because the charged-particle motion in this class of rotating backgrounds is integrable, the instability of an equilibrium orbit is a local property and not a sign of chaos, whereas genuine chaos appears only when integrability is broken and then remains well within the chaos bound. Together with an exact treatment of frame dragging, this distinction provides a consistent framework for testing the chaos bound with classical probes in modified gravity.


\section*{Data Availability Statement}

No new data were created or analyzed in this study. Data sharing is not applicable to this article.

\section*{Conflict of Interest}

The authors declare that there is no conflict of interest.

\bibliographystyle{apsrev4-1}
\bibliography{ChaosGB} 

\begin{thebibliography}{61}%
\makeatletter
\providecommand \@ifxundefined [1]{%
 \@ifx{#1\undefined}
}%
\providecommand \@ifnum [1]{%
 \ifnum #1\expandafter \@firstoftwo
 \else \expandafter \@secondoftwo
 \fi
}%
\providecommand \@ifx [1]{%
 \ifx #1\expandafter \@firstoftwo
 \else \expandafter \@secondoftwo
 \fi
}%
\providecommand \natexlab [1]{#1}%
\providecommand \enquote  [1]{``#1''}%
\providecommand \bibnamefont  [1]{#1}%
\providecommand \bibfnamefont [1]{#1}%
\providecommand \citenamefont [1]{#1}%
\providecommand \href@noop [0]{\@secondoftwo}%
\providecommand \href [0]{\begingroup \@sanitize@url \@href}%
\providecommand \@href[1]{\@@startlink{#1}\@@href}%
\providecommand \@@href[1]{\endgroup#1\@@endlink}%
\providecommand \@sanitize@url [0]{\catcode `\\12\catcode `\$12\catcode `\&12\catcode `\#12\catcode `\^12\catcode `\_12\catcode `\%12\relax}%
\providecommand \@@startlink[1]{}%
\providecommand \@@endlink[0]{}%
\providecommand \url  [0]{\begingroup\@sanitize@url \@url }%
\providecommand \@url [1]{\endgroup\@href {#1}{\urlprefix }}%
\providecommand \urlprefix  [0]{URL }%
\providecommand \Eprint [0]{\href }%
\providecommand \doibase [0]{http://dx.doi.org/}%
\providecommand \selectlanguage [0]{\@gobble}%
\providecommand \bibinfo  [0]{\@secondoftwo}%
\providecommand \bibfield  [0]{\@secondoftwo}%
\providecommand \translation [1]{[#1]}%
\providecommand \BibitemOpen [0]{}%
\providecommand \bibitemStop [0]{}%
\providecommand \bibitemNoStop [0]{.\EOS\space}%
\providecommand \EOS [0]{\spacefactor3000\relax}%
\providecommand \BibitemShut  [1]{\csname bibitem#1\endcsname}%
\let\auto@bib@innerbib\@empty
\bibitem [{\citenamefont {Zwiebach}(1985)}]{Zwiebach1985}%
  \BibitemOpen
  \bibfield  {author} {\bibinfo {author} {\bibfnamefont {B.}~\bibnamefont {Zwiebach}},\ }\href {\doibase 10.1016/0370-2693(85)91616-8} {\bibfield  {journal} {\bibinfo  {journal} {Physics Letters B}\ }\textbf {\bibinfo {volume} {156}},\ \bibinfo {pages} {315} (\bibinfo {year} {1985})}\BibitemShut {NoStop}%
\bibitem [{\citenamefont {Garraffo}\ and\ \citenamefont {Giribet}(2008)}]{Garraffo2008}%
  \BibitemOpen
  \bibfield  {author} {\bibinfo {author} {\bibfnamefont {C.}~\bibnamefont {Garraffo}}\ and\ \bibinfo {author} {\bibfnamefont {G.}~\bibnamefont {Giribet}},\ }\href {\doibase 10.1142/S0217732308027567} {\bibfield  {journal} {\bibinfo  {journal} {Modern Physics Letters A}\ }\textbf {\bibinfo {volume} {23}},\ \bibinfo {pages} {1801} (\bibinfo {year} {2008})}\BibitemShut {NoStop}%
\bibitem [{\citenamefont {Glavan}\ and\ \citenamefont {Lin}(2020)}]{Glavan2020}%
  \BibitemOpen
  \bibfield  {author} {\bibinfo {author} {\bibfnamefont {D.}~\bibnamefont {Glavan}}\ and\ \bibinfo {author} {\bibfnamefont {C.}~\bibnamefont {Lin}},\ }\href {\doibase 10.1103/PhysRevLett.124.081301} {\bibfield  {journal} {\bibinfo  {journal} {Physical Review Letters}\ }\textbf {\bibinfo {volume} {124}},\ \bibinfo {pages} {081301} (\bibinfo {year} {2020})}\BibitemShut {NoStop}%
\bibitem [{\citenamefont {Cornish}\ and\ \citenamefont {Frankel}(1997)}]{Cornish:1996ri}%
  \BibitemOpen
  \bibfield  {author} {\bibinfo {author} {\bibfnamefont {N.~J.}\ \bibnamefont {Cornish}}\ and\ \bibinfo {author} {\bibfnamefont {N.~E.}\ \bibnamefont {Frankel}},\ }\href@noop {} {\bibfield  {journal} {\bibinfo  {journal} {Phys. Rev. D}\ }\textbf {\bibinfo {volume} {56}},\ \bibinfo {pages} {1903} (\bibinfo {year} {1997})}\BibitemShut {NoStop}%
\bibitem [{\citenamefont {Bombelli}\ \emph {et~al.}(1991)\citenamefont {Bombelli}, \citenamefont {Calzetta},\ and\ \citenamefont {El~Hasi}}]{Bombelli:1991eg}%
  \BibitemOpen
  \bibfield  {author} {\bibinfo {author} {\bibfnamefont {L.}~\bibnamefont {Bombelli}}, \bibinfo {author} {\bibfnamefont {E.}~\bibnamefont {Calzetta}}, \ and\ \bibinfo {author} {\bibfnamefont {C.}~\bibnamefont {El~Hasi}},\ }\href@noop {} {\bibfield  {journal} {\bibinfo  {journal} {Phys. Rev. D}\ }\textbf {\bibinfo {volume} {43}},\ \bibinfo {pages} {2079} (\bibinfo {year} {1991})}\BibitemShut {NoStop}%
\bibitem [{\citenamefont {Zayas}\ and\ \citenamefont {Terrero-Escalante}(2010)}]{Zayas:2010fs}%
  \BibitemOpen
  \bibfield  {author} {\bibinfo {author} {\bibfnamefont {L.~A.~P.}\ \bibnamefont {Zayas}}\ and\ \bibinfo {author} {\bibfnamefont {C.~A.}\ \bibnamefont {Terrero-Escalante}},\ }\href@noop {} {\bibfield  {journal} {\bibinfo  {journal} {JHEP}\ }\textbf {\bibinfo {volume} {09}},\ \bibinfo {pages} {094} (\bibinfo {year} {2010})}\BibitemShut {NoStop}%
\bibitem [{\citenamefont {Hashimoto}\ and\ \citenamefont {Tanahashi}(2017)}]{Hashimoto2017Universality}%
  \BibitemOpen
  \bibfield  {author} {\bibinfo {author} {\bibfnamefont {K.}~\bibnamefont {Hashimoto}}\ and\ \bibinfo {author} {\bibfnamefont {N.}~\bibnamefont {Tanahashi}},\ }\href {\doibase 10.1103/PhysRevD.95.024007} {\bibfield  {journal} {\bibinfo  {journal} {Physical Review D}\ }\textbf {\bibinfo {volume} {95}},\ \bibinfo {pages} {024007} (\bibinfo {year} {2017})}\BibitemShut {NoStop}%
\bibitem [{\citenamefont {Hashimoto}\ \emph {et~al.}(2018)\citenamefont {Hashimoto}, \citenamefont {Murata},\ and\ \citenamefont {Tanahashi}}]{Hashimoto:2018fkb}%
  \BibitemOpen
  \bibfield  {author} {\bibinfo {author} {\bibfnamefont {K.}~\bibnamefont {Hashimoto}}, \bibinfo {author} {\bibfnamefont {K.}~\bibnamefont {Murata}}, \ and\ \bibinfo {author} {\bibfnamefont {N.}~\bibnamefont {Tanahashi}},\ }\href {\doibase 10.1103/PhysRevD.98.086007} {\bibfield  {journal} {\bibinfo  {journal} {Phys. Rev. D}\ }\textbf {\bibinfo {volume} {98}},\ \bibinfo {pages} {086007} (\bibinfo {year} {2018})},\ \Eprint {http://arxiv.org/abs/1803.06756} {arXiv:1803.06756 [hep-th]} \BibitemShut {NoStop}%
\bibitem [{\citenamefont {Maldacena}\ \emph {et~al.}(2016)\citenamefont {Maldacena}, \citenamefont {Shenker},\ and\ \citenamefont {Stanford}}]{Maldacena:2015waa}%
  \BibitemOpen
  \bibfield  {author} {\bibinfo {author} {\bibfnamefont {J.}~\bibnamefont {Maldacena}}, \bibinfo {author} {\bibfnamefont {S.~H.}\ \bibnamefont {Shenker}}, \ and\ \bibinfo {author} {\bibfnamefont {D.}~\bibnamefont {Stanford}},\ }\href {\doibase 10.1007/JHEP08(2016)106} {\bibfield  {journal} {\bibinfo  {journal} {JHEP}\ }\textbf {\bibinfo {volume} {08}},\ \bibinfo {pages} {106} (\bibinfo {year} {2016})},\ \Eprint {http://arxiv.org/abs/1503.01409} {arXiv:1503.01409 [hep-th]} \BibitemShut {NoStop}%
\bibitem [{\citenamefont {Lei}\ \emph {et~al.}(2021{\natexlab{a}})\citenamefont {Lei}, \citenamefont {Ge},\ and\ \citenamefont {Ran}}]{Lei2021Chaos}%
  \BibitemOpen
  \bibfield  {author} {\bibinfo {author} {\bibfnamefont {Y.-Q.}\ \bibnamefont {Lei}}, \bibinfo {author} {\bibfnamefont {X.-H.}\ \bibnamefont {Ge}}, \ and\ \bibinfo {author} {\bibfnamefont {C.}~\bibnamefont {Ran}},\ }\href {\doibase 10.1103/PhysRevD.104.046020} {\bibfield  {journal} {\bibinfo  {journal} {Physical Review D}\ }\textbf {\bibinfo {volume} {104}},\ \bibinfo {pages} {046020} (\bibinfo {year} {2021}{\natexlab{a}})}\BibitemShut {NoStop}%
\bibitem [{\citenamefont {Bokhari}\ \emph {et~al.}(2020)\citenamefont {Bokhari}, \citenamefont {Rayimbaev},\ and\ \citenamefont {Ahmedov}}]{Bokhari2020TestParticles}%
  \BibitemOpen
  \bibfield  {author} {\bibinfo {author} {\bibfnamefont {A.~H.}\ \bibnamefont {Bokhari}}, \bibinfo {author} {\bibfnamefont {J.}~\bibnamefont {Rayimbaev}}, \ and\ \bibinfo {author} {\bibfnamefont {B.}~\bibnamefont {Ahmedov}},\ }\href {\doibase 10.1103/PhysRevD.102.124078} {\bibfield  {journal} {\bibinfo  {journal} {Physical Review D}\ }\textbf {\bibinfo {volume} {102}},\ \bibinfo {pages} {124078} (\bibinfo {year} {2020})}\BibitemShut {NoStop}%
\bibitem [{\citenamefont {Guo}\ \emph {et~al.}(2021)\citenamefont {Guo}, \citenamefont {Liang}, \citenamefont {Mu}, \citenamefont {Wang},\ and\ \citenamefont {Yang}}]{Guo2021Minimal}%
  \BibitemOpen
  \bibfield  {author} {\bibinfo {author} {\bibfnamefont {X.}~\bibnamefont {Guo}}, \bibinfo {author} {\bibfnamefont {K.}~\bibnamefont {Liang}}, \bibinfo {author} {\bibfnamefont {B.}~\bibnamefont {Mu}}, \bibinfo {author} {\bibfnamefont {P.}~\bibnamefont {Wang}}, \ and\ \bibinfo {author} {\bibfnamefont {M.}~\bibnamefont {Yang}},\ }\href {\doibase 10.1088/1674-1137/abc4d6} {\bibfield  {journal} {\bibinfo  {journal} {Chinese Physics C}\ }\textbf {\bibinfo {volume} {45}},\ \bibinfo {pages} {023115} (\bibinfo {year} {2021})}\BibitemShut {NoStop}%
\bibitem [{\citenamefont {Chen}\ and\ \citenamefont {Gao}(2022)}]{ChenGao2022}%
  \BibitemOpen
  \bibfield  {author} {\bibinfo {author} {\bibfnamefont {D.}~\bibnamefont {Chen}}\ and\ \bibinfo {author} {\bibfnamefont {C.}~\bibnamefont {Gao}},\ }\href {\doibase 10.1088/1367-2630/ac9d5f} {\bibfield  {journal} {\bibinfo  {journal} {New Journal of Physics}\ }\textbf {\bibinfo {volume} {24}},\ \bibinfo {pages} {123014} (\bibinfo {year} {2022})}\BibitemShut {NoStop}%
\bibitem [{\citenamefont {Yu}\ \emph {et~al.}(2023)\citenamefont {Yu}, \citenamefont {Chen}, \citenamefont {Mu},\ and\ \citenamefont {He}}]{Yu2023}%
  \BibitemOpen
  \bibfield  {author} {\bibinfo {author} {\bibfnamefont {C.}~\bibnamefont {Yu}}, \bibinfo {author} {\bibfnamefont {D.}~\bibnamefont {Chen}}, \bibinfo {author} {\bibfnamefont {B.}~\bibnamefont {Mu}}, \ and\ \bibinfo {author} {\bibfnamefont {Y.}~\bibnamefont {He}},\ }\href {\doibase 10.1016/j.nuclphysb.2023.116093} {\bibfield  {journal} {\bibinfo  {journal} {Nuclear Physics B}\ }\textbf {\bibinfo {volume} {987}},\ \bibinfo {pages} {116093} (\bibinfo {year} {2023})}\BibitemShut {NoStop}%
\bibitem [{\citenamefont {Gao}\ \emph {et~al.}(2026)\citenamefont {Gao}, \citenamefont {Yang}, \citenamefont {Chong},\ and\ \citenamefont {Chen}}]{Gao2026}%
  \BibitemOpen
  \bibfield  {author} {\bibinfo {author} {\bibfnamefont {C.}~\bibnamefont {Gao}}, \bibinfo {author} {\bibfnamefont {C.}~\bibnamefont {Yang}}, \bibinfo {author} {\bibfnamefont {T.}~\bibnamefont {Chong}}, \ and\ \bibinfo {author} {\bibfnamefont {D.}~\bibnamefont {Chen}},\ }\href {\doibase 10.1088/1674-1137/ae210d} {\bibfield  {journal} {\bibinfo  {journal} {Chinese Physics C}\ }\textbf {\bibinfo {volume} {50}},\ \bibinfo {pages} {035101} (\bibinfo {year} {2026})}\BibitemShut {NoStop}%
\bibitem [{\citenamefont {Yu}\ \emph {et~al.}(2022)\citenamefont {Yu}, \citenamefont {Chen},\ and\ \citenamefont {Gao}}]{YuChenGao2022}%
  \BibitemOpen
  \bibfield  {author} {\bibinfo {author} {\bibfnamefont {C.}~\bibnamefont {Yu}}, \bibinfo {author} {\bibfnamefont {D.}~\bibnamefont {Chen}}, \ and\ \bibinfo {author} {\bibfnamefont {C.}~\bibnamefont {Gao}},\ }\href {\doibase 10.1088/1674-1137/ac8f4e} {\bibfield  {journal} {\bibinfo  {journal} {Chinese Physics C}\ }\textbf {\bibinfo {volume} {46}},\ \bibinfo {pages} {125106} (\bibinfo {year} {2022})}\BibitemShut {NoStop}%
\bibitem [{\citenamefont {Wang}\ and\ \citenamefont {Chen}(2023)}]{WangChen2023}%
  \BibitemOpen
  \bibfield  {author} {\bibinfo {author} {\bibfnamefont {Z.}~\bibnamefont {Wang}}\ and\ \bibinfo {author} {\bibfnamefont {D.}~\bibnamefont {Chen}},\ }\href {\doibase 10.1016/j.nuclphysb.2023.116212} {\bibfield  {journal} {\bibinfo  {journal} {Nuclear Physics B}\ }\textbf {\bibinfo {volume} {991}},\ \bibinfo {pages} {116212} (\bibinfo {year} {2023})}\BibitemShut {NoStop}%
\bibitem [{\citenamefont {Audin}\ \emph {et~al.}(2012)\citenamefont {Audin}, \citenamefont {Da~Silva},\ and\ \citenamefont {Lerman}}]{audin2012symplectic}%
  \BibitemOpen
  \bibfield  {author} {\bibinfo {author} {\bibfnamefont {M.}~\bibnamefont {Audin}}, \bibinfo {author} {\bibfnamefont {A.~C.}\ \bibnamefont {Da~Silva}}, \ and\ \bibinfo {author} {\bibfnamefont {E.}~\bibnamefont {Lerman}},\ }\href@noop {} {\emph {\bibinfo {title} {Symplectic geometry of integrable Hamiltonian systems}}}\ (\bibinfo  {publisher} {Birkh{\"a}user},\ \bibinfo {year} {2012})\BibitemShut {NoStop}%
\bibitem [{\citenamefont {Wang}\ \emph {et~al.}(2021)\citenamefont {Wang}, \citenamefont {Sun}, \citenamefont {Liu},\ and\ \citenamefont {Wu}}]{Wang:2021gja}%
  \BibitemOpen
  \bibfield  {author} {\bibinfo {author} {\bibfnamefont {Y.}~\bibnamefont {Wang}}, \bibinfo {author} {\bibfnamefont {W.}~\bibnamefont {Sun}}, \bibinfo {author} {\bibfnamefont {F.}~\bibnamefont {Liu}}, \ and\ \bibinfo {author} {\bibfnamefont {X.}~\bibnamefont {Wu}},\ }\href {\doibase 10.3847/1538-4357/abcb8d} {\bibfield  {journal} {\bibinfo  {journal} {Astrophys. J.}\ }\textbf {\bibinfo {volume} {907}},\ \bibinfo {pages} {66} (\bibinfo {year} {2021})},\ \Eprint {http://arxiv.org/abs/2102.00373} {arXiv:2102.00373 [gr-qc]} \BibitemShut {NoStop}%
\bibitem [{\citenamefont {Chen}\ \emph {et~al.}(2016)\citenamefont {Chen}, \citenamefont {Wang},\ and\ \citenamefont {Jing}}]{Chen:2016tmr}%
  \BibitemOpen
  \bibfield  {author} {\bibinfo {author} {\bibfnamefont {S.}~\bibnamefont {Chen}}, \bibinfo {author} {\bibfnamefont {M.}~\bibnamefont {Wang}}, \ and\ \bibinfo {author} {\bibfnamefont {J.}~\bibnamefont {Jing}},\ }\href {\doibase 10.1007/JHEP09(2016)082} {\bibfield  {journal} {\bibinfo  {journal} {JHEP}\ }\textbf {\bibinfo {volume} {09}},\ \bibinfo {pages} {082} (\bibinfo {year} {2016})},\ \Eprint {http://arxiv.org/abs/1604.02785} {arXiv:1604.02785 [gr-qc]} \BibitemShut {NoStop}%
\bibitem [{\citenamefont {Lei}\ \emph {et~al.}(2021{\natexlab{b}})\citenamefont {Lei}, \citenamefont {Ge},\ and\ \citenamefont {Ran}}]{Lei:2020clg}%
  \BibitemOpen
  \bibfield  {author} {\bibinfo {author} {\bibfnamefont {Y.-Q.}\ \bibnamefont {Lei}}, \bibinfo {author} {\bibfnamefont {X.-H.}\ \bibnamefont {Ge}}, \ and\ \bibinfo {author} {\bibfnamefont {C.}~\bibnamefont {Ran}},\ }\href {\doibase 10.1103/PhysRevD.104.046020} {\bibfield  {journal} {\bibinfo  {journal} {Phys. Rev. D}\ }\textbf {\bibinfo {volume} {104}},\ \bibinfo {pages} {046020} (\bibinfo {year} {2021}{\natexlab{b}})},\ \Eprint {http://arxiv.org/abs/2008.01384} {arXiv:2008.01384 [hep-th]} \BibitemShut {NoStop}%
\bibitem [{\citenamefont {Basu}\ \emph {et~al.}(2012)\citenamefont {Basu}, \citenamefont {Das}, \citenamefont {Ghosh},\ and\ \citenamefont {Pando~Zayas}}]{Basu:2012ae}%
  \BibitemOpen
  \bibfield  {author} {\bibinfo {author} {\bibfnamefont {P.}~\bibnamefont {Basu}}, \bibinfo {author} {\bibfnamefont {D.}~\bibnamefont {Das}}, \bibinfo {author} {\bibfnamefont {A.}~\bibnamefont {Ghosh}}, \ and\ \bibinfo {author} {\bibfnamefont {L.~A.}\ \bibnamefont {Pando~Zayas}},\ }\href {\doibase 10.1007/JHEP05(2012)077} {\bibfield  {journal} {\bibinfo  {journal} {JHEP}\ }\textbf {\bibinfo {volume} {05}},\ \bibinfo {pages} {077} (\bibinfo {year} {2012})},\ \Eprint {http://arxiv.org/abs/1201.5634} {arXiv:1201.5634 [hep-th]} \BibitemShut {NoStop}%
\bibitem [{\citenamefont {Deich}\ \emph {et~al.}(2022)\citenamefont {Deich}, \citenamefont {C{\'a}rdenas-Avenda{\~n}o},\ and\ \citenamefont {Yunes}}]{Deich:2022vna}%
  \BibitemOpen
  \bibfield  {author} {\bibinfo {author} {\bibfnamefont {A.}~\bibnamefont {Deich}}, \bibinfo {author} {\bibfnamefont {A.}~\bibnamefont {C{\'a}rdenas-Avenda{\~n}o}}, \ and\ \bibinfo {author} {\bibfnamefont {N.}~\bibnamefont {Yunes}},\ }\href {\doibase 10.1103/PhysRevD.106.024040} {\bibfield  {journal} {\bibinfo  {journal} {Phys. Rev. D}\ }\textbf {\bibinfo {volume} {106}},\ \bibinfo {pages} {024040} (\bibinfo {year} {2022})},\ \Eprint {http://arxiv.org/abs/2203.00524} {arXiv:2203.00524 [gr-qc]} \BibitemShut {NoStop}%
\bibitem [{\citenamefont {Pikovsky}\ and\ \citenamefont {Politi}(1945)}]{pikovsky1945lyapunov}%
  \BibitemOpen
  \bibfield  {author} {\bibinfo {author} {\bibfnamefont {A.}~\bibnamefont {Pikovsky}}\ and\ \bibinfo {author} {\bibfnamefont {A.}~\bibnamefont {Politi}},\ }\href@noop {} {\bibfield  {journal} {\bibinfo  {journal} {American history}\ }\textbf {\bibinfo {volume} {1861}} (\bibinfo {year} {1945})}\BibitemShut {NoStop}%
\bibitem [{\citenamefont {Sandri}(1996)}]{sandri1996numerical}%
  \BibitemOpen
  \bibfield  {author} {\bibinfo {author} {\bibfnamefont {M.}~\bibnamefont {Sandri}},\ }\href@noop {} {\bibfield  {journal} {\bibinfo  {journal} {Mathematica Journal}\ }\textbf {\bibinfo {volume} {6}},\ \bibinfo {pages} {78} (\bibinfo {year} {1996})}\BibitemShut {NoStop}%
\bibitem [{\citenamefont {El-Nabulsi}\ \emph {et~al.}(2025)\citenamefont {El-Nabulsi}, \citenamefont {Thangaraj}, \citenamefont {Valarmathi},\ and\ \citenamefont {Anukool}}]{el2025chaotic}%
  \BibitemOpen
  \bibfield  {author} {\bibinfo {author} {\bibfnamefont {R.~A.}\ \bibnamefont {El-Nabulsi}}, \bibinfo {author} {\bibfnamefont {C.}~\bibnamefont {Thangaraj}}, \bibinfo {author} {\bibfnamefont {R.}~\bibnamefont {Valarmathi}}, \ and\ \bibinfo {author} {\bibfnamefont {W.}~\bibnamefont {Anukool}},\ }\href@noop {} {\bibfield  {journal} {\bibinfo  {journal} {Chaos, Solitons \& Fractals}\ }\textbf {\bibinfo {volume} {200}},\ \bibinfo {pages} {116974} (\bibinfo {year} {2025})}\BibitemShut {NoStop}%
\bibitem [{\citenamefont {El-Nabulsi}\ \emph {et~al.}(2026{\natexlab{a}})\citenamefont {El-Nabulsi}, \citenamefont {Anukool}, \citenamefont {Valarmathi},\ and\ \citenamefont {Thangaraj}}]{el2026extended}%
  \BibitemOpen
  \bibfield  {author} {\bibinfo {author} {\bibfnamefont {R.~A.}\ \bibnamefont {El-Nabulsi}}, \bibinfo {author} {\bibfnamefont {W.}~\bibnamefont {Anukool}}, \bibinfo {author} {\bibfnamefont {R.}~\bibnamefont {Valarmathi}}, \ and\ \bibinfo {author} {\bibfnamefont {C.}~\bibnamefont {Thangaraj}},\ }\href@noop {} {\bibfield  {journal} {\bibinfo  {journal} {Advances in Space Research}\ } (\bibinfo {year} {2026}{\natexlab{a}})}\BibitemShut {NoStop}%
\bibitem [{\citenamefont {El-Nabulsi}\ \emph {et~al.}(2026{\natexlab{b}})\citenamefont {El-Nabulsi}, \citenamefont {Anukool}, \citenamefont {Thangaraj},\ and\ \citenamefont {Valarmathi}}]{el2026stability}%
  \BibitemOpen
  \bibfield  {author} {\bibinfo {author} {\bibfnamefont {R.~A.}\ \bibnamefont {El-Nabulsi}}, \bibinfo {author} {\bibfnamefont {W.}~\bibnamefont {Anukool}}, \bibinfo {author} {\bibfnamefont {C.}~\bibnamefont {Thangaraj}}, \ and\ \bibinfo {author} {\bibfnamefont {R.}~\bibnamefont {Valarmathi}},\ }\href@noop {} {\bibfield  {journal} {\bibinfo  {journal} {Chaos, Solitons \& Fractals}\ }\textbf {\bibinfo {volume} {202}},\ \bibinfo {pages} {117410} (\bibinfo {year} {2026}{\natexlab{b}})}\BibitemShut {NoStop}%
\bibitem [{\citenamefont {El-Nabulsi}\ and\ \citenamefont {Anukool}(2026)}]{El-Nabulsi:2026pls}%
  \BibitemOpen
  \bibfield  {author} {\bibinfo {author} {\bibfnamefont {R.~A.}\ \bibnamefont {El-Nabulsi}}\ and\ \bibinfo {author} {\bibfnamefont {W.}~\bibnamefont {Anukool}},\ }\href {\doibase 10.1142/s0217732326500847} {\bibfield  {journal} {\bibinfo  {journal} {Mod. Phys. Lett. A}\ }\textbf {\bibinfo {volume} {41}},\ \bibinfo {pages} {2650084} (\bibinfo {year} {2026})}\BibitemShut {NoStop}%
\bibitem [{\citenamefont {Szumi{\'n}ski}\ \emph {et~al.}(2024)\citenamefont {Szumi{\'n}ski}, \citenamefont {Przybylska},\ and\ \citenamefont {Maciejewski}}]{NLD2023curved}%
  \BibitemOpen
  \bibfield  {author} {\bibinfo {author} {\bibfnamefont {W.}~\bibnamefont {Szumi{\'n}ski}}, \bibinfo {author} {\bibfnamefont {M.}~\bibnamefont {Przybylska}}, \ and\ \bibinfo {author} {\bibfnamefont {A.~J.}\ \bibnamefont {Maciejewski}},\ }\href {\doibase 10.1007/s11071-023-09224-w} {\bibfield  {journal} {\bibinfo  {journal} {Nonlinear Dyn.}\ }\textbf {\bibinfo {volume} {112}},\ \bibinfo {pages} {4879} (\bibinfo {year} {2024})}\BibitemShut {NoStop}%
\bibitem [{\citenamefont {Szumi{\'n}ski}\ and\ \citenamefont {Przybylska}(2025)}]{Szuminski:2025nld}%
  \BibitemOpen
  \bibfield  {author} {\bibinfo {author} {\bibfnamefont {W.}~\bibnamefont {Szumi{\'n}ski}}\ and\ \bibinfo {author} {\bibfnamefont {M.}~\bibnamefont {Przybylska}},\ }\href {\doibase 10.1007/s11071-025-11554-w} {\bibfield  {journal} {\bibinfo  {journal} {Nonlinear Dyn.}\ }\textbf {\bibinfo {volume} {113}},\ \bibinfo {pages} {30057} (\bibinfo {year} {2025})}\BibitemShut {NoStop}%
\bibitem [{\citenamefont {Carter}(1968)}]{Carter:1968rr}%
  \BibitemOpen
  \bibfield  {author} {\bibinfo {author} {\bibfnamefont {B.}~\bibnamefont {Carter}},\ }\href {\doibase 10.1103/PhysRev.174.1559} {\bibfield  {journal} {\bibinfo  {journal} {Phys. Rev.}\ }\textbf {\bibinfo {volume} {174}},\ \bibinfo {pages} {1559} (\bibinfo {year} {1968})}\BibitemShut {NoStop}%
\bibitem [{\citenamefont {Dalui}\ \emph {et~al.}(2019{\natexlab{a}})\citenamefont {Dalui}, \citenamefont {Majhi},\ and\ \citenamefont {Mishra}}]{Dalui2019Presence}%
  \BibitemOpen
  \bibfield  {author} {\bibinfo {author} {\bibfnamefont {S.}~\bibnamefont {Dalui}}, \bibinfo {author} {\bibfnamefont {B.~R.}\ \bibnamefont {Majhi}}, \ and\ \bibinfo {author} {\bibfnamefont {P.}~\bibnamefont {Mishra}},\ }\href {\doibase 10.1016/j.physletb.2018.11.050} {\bibfield  {journal} {\bibinfo  {journal} {Physics Letters B}\ }\textbf {\bibinfo {volume} {788}},\ \bibinfo {pages} {486} (\bibinfo {year} {2019}{\natexlab{a}})}\BibitemShut {NoStop}%
\bibitem [{\citenamefont {Dalui}\ \emph {et~al.}(2020)\citenamefont {Dalui}, \citenamefont {Majhi},\ and\ \citenamefont {Mishra}}]{Dalui2020Induction}%
  \BibitemOpen
  \bibfield  {author} {\bibinfo {author} {\bibfnamefont {S.}~\bibnamefont {Dalui}}, \bibinfo {author} {\bibfnamefont {B.~R.}\ \bibnamefont {Majhi}}, \ and\ \bibinfo {author} {\bibfnamefont {P.}~\bibnamefont {Mishra}},\ }\href {\doibase 10.1142/S0217751X20500813} {\bibfield  {journal} {\bibinfo  {journal} {International Journal of Modern Physics A}\ }\textbf {\bibinfo {volume} {35}},\ \bibinfo {pages} {2050081} (\bibinfo {year} {2020})}\BibitemShut {NoStop}%
\bibitem [{\citenamefont {Chen}\ \emph {et~al.}(2019)\citenamefont {Chen}, \citenamefont {Li},\ and\ \citenamefont {Zhang}}]{Chen2019BornInfeld}%
  \BibitemOpen
  \bibfield  {author} {\bibinfo {author} {\bibfnamefont {Y.}~\bibnamefont {Chen}}, \bibinfo {author} {\bibfnamefont {H.}~\bibnamefont {Li}}, \ and\ \bibinfo {author} {\bibfnamefont {J.}~\bibnamefont {Zhang}},\ }\href {\doibase 10.1007/s10714-019-2606-7} {\bibfield  {journal} {\bibinfo  {journal} {General Relativity and Gravitation}\ }\textbf {\bibinfo {volume} {51}},\ \bibinfo {pages} {134} (\bibinfo {year} {2019})}\BibitemShut {NoStop}%
\bibitem [{\citenamefont {Kala}\ \emph {et~al.}(2026{\natexlab{a}})\citenamefont {Kala}, \citenamefont {Singh}, \citenamefont {Yousaf}, \citenamefont {Nandan}, \citenamefont {Atamurotov},\ and\ \citenamefont {Mustafa}}]{Kala:2026zuf}%
  \BibitemOpen
  \bibfield  {author} {\bibinfo {author} {\bibfnamefont {S.}~\bibnamefont {Kala}}, \bibinfo {author} {\bibfnamefont {P.}~\bibnamefont {Singh}}, \bibinfo {author} {\bibfnamefont {M.}~\bibnamefont {Yousaf}}, \bibinfo {author} {\bibfnamefont {H.}~\bibnamefont {Nandan}}, \bibinfo {author} {\bibfnamefont {F.}~\bibnamefont {Atamurotov}}, \ and\ \bibinfo {author} {\bibfnamefont {G.}~\bibnamefont {Mustafa}},\ }\href {\doibase 10.1016/j.chaos.2026.119360} {\bibfield  {journal} {\bibinfo  {journal} {Chaos Solitons Fractals}\ }\textbf {\bibinfo {volume} {213}},\ \bibinfo {pages} {119360} (\bibinfo {year} {2026}{\natexlab{a}})}\BibitemShut {NoStop}%
\bibitem [{\citenamefont {Kala}\ \emph {et~al.}(2026{\natexlab{b}})\citenamefont {Kala}, \citenamefont {Yousaf}, \citenamefont {Yuan},\ and\ \citenamefont {Mustafa}}]{Kala:20267}%
  \BibitemOpen
  \bibfield  {author} {\bibinfo {author} {\bibfnamefont {S.}~\bibnamefont {Kala}}, \bibinfo {author} {\bibfnamefont {M.}~\bibnamefont {Yousaf}}, \bibinfo {author} {\bibfnamefont {C.}~\bibnamefont {Yuan}}, \ and\ \bibinfo {author} {\bibfnamefont {G.}~\bibnamefont {Mustafa}},\ }\href@noop {} {\bibfield  {journal} {\bibinfo  {journal} {Accepted for publication in Chaos Solitons Fractals}\ } (\bibinfo {year} {2026}{\natexlab{b}})}\BibitemShut {NoStop}%
\bibitem [{\citenamefont {Li}\ \emph {et~al.}(2025)\citenamefont {Li}, \citenamefont {Ma},\ and\ \citenamefont {Xu}}]{Li2025StringChaos}%
  \BibitemOpen
  \bibfield  {author} {\bibinfo {author} {\bibfnamefont {K.}~\bibnamefont {Li}}, \bibinfo {author} {\bibfnamefont {D.-Z.}\ \bibnamefont {Ma}}, \ and\ \bibinfo {author} {\bibfnamefont {Z.-M.}\ \bibnamefont {Xu}},\ }\href {\doibase 10.1016/j.physletb.2024.139164} {\bibfield  {journal} {\bibinfo  {journal} {Physics Letters B}\ }\textbf {\bibinfo {volume} {860}},\ \bibinfo {pages} {139164} (\bibinfo {year} {2025})}\BibitemShut {NoStop}%
\bibitem [{\citenamefont {Chen}\ and\ \citenamefont {Gao}(2023)}]{chen2023chaos}%
  \BibitemOpen
  \bibfield  {author} {\bibinfo {author} {\bibfnamefont {D.}~\bibnamefont {Chen}}\ and\ \bibinfo {author} {\bibfnamefont {C.}~\bibnamefont {Gao}},\ }\href {\doibase 10.1088/1674-1137/ac9fb9} {\bibfield  {journal} {\bibinfo  {journal} {Chinese Physics C}\ }\textbf {\bibinfo {volume} {47}},\ \bibinfo {pages} {015108} (\bibinfo {year} {2023})}\BibitemShut {NoStop}%
\bibitem [{\citenamefont {Nosirov}\ \emph {et~al.}(2024)\citenamefont {Nosirov}, \citenamefont {Atamurotov}, \citenamefont {Rakhimova}, \citenamefont {Abdujabbarov},\ and\ \citenamefont {Ghosh}}]{nosirov2024particle}%
  \BibitemOpen
  \bibfield  {author} {\bibinfo {author} {\bibfnamefont {A.}~\bibnamefont {Nosirov}}, \bibinfo {author} {\bibfnamefont {F.}~\bibnamefont {Atamurotov}}, \bibinfo {author} {\bibfnamefont {G.}~\bibnamefont {Rakhimova}}, \bibinfo {author} {\bibfnamefont {A.}~\bibnamefont {Abdujabbarov}}, \ and\ \bibinfo {author} {\bibfnamefont {S.~G.}\ \bibnamefont {Ghosh}},\ }\href@noop {} {\bibfield  {journal} {\bibinfo  {journal} {Nuclear Physics B}\ }\textbf {\bibinfo {volume} {1005}},\ \bibinfo {pages} {116583} (\bibinfo {year} {2024})}\BibitemShut {NoStop}%
\bibitem [{\citenamefont {Li}\ \emph {et~al.}(2022)\citenamefont {Li}, \citenamefont {Duan},\ and\ \citenamefont {Jia}}]{li2022deflection}%
  \BibitemOpen
  \bibfield  {author} {\bibinfo {author} {\bibfnamefont {Z.}~\bibnamefont {Li}}, \bibinfo {author} {\bibfnamefont {Y.}~\bibnamefont {Duan}}, \ and\ \bibinfo {author} {\bibfnamefont {J.}~\bibnamefont {Jia}},\ }\href@noop {} {\bibfield  {journal} {\bibinfo  {journal} {Classical and Quantum Gravity}\ }\textbf {\bibinfo {volume} {39}},\ \bibinfo {pages} {015002} (\bibinfo {year} {2022})}\BibitemShut {NoStop}%
\bibitem [{\citenamefont {Papnoi}\ and\ \citenamefont {Atamurotov}(2022)}]{papnoi2022rotating}%
  \BibitemOpen
  \bibfield  {author} {\bibinfo {author} {\bibfnamefont {U.}~\bibnamefont {Papnoi}}\ and\ \bibinfo {author} {\bibfnamefont {F.}~\bibnamefont {Atamurotov}},\ }\href@noop {} {\bibfield  {journal} {\bibinfo  {journal} {Physics of the Dark Universe}\ }\textbf {\bibinfo {volume} {35}},\ \bibinfo {pages} {100916} (\bibinfo {year} {2022})}\BibitemShut {NoStop}%
\bibitem [{\citenamefont {Hansen}\ and\ \citenamefont {Yunes}(2013)}]{Hansen:2013owa}%
  \BibitemOpen
  \bibfield  {author} {\bibinfo {author} {\bibfnamefont {D.}~\bibnamefont {Hansen}}\ and\ \bibinfo {author} {\bibfnamefont {N.}~\bibnamefont {Yunes}},\ }\href {\doibase 10.1103/PhysRevD.88.104020} {\bibfield  {journal} {\bibinfo  {journal} {Phys. Rev. D}\ }\textbf {\bibinfo {volume} {88}},\ \bibinfo {pages} {104020} (\bibinfo {year} {2013})},\ \Eprint {http://arxiv.org/abs/1308.6631} {arXiv:1308.6631 [gr-qc]} \BibitemShut {NoStop}%
\bibitem [{\citenamefont {Fayyaz}\ \emph {et~al.}(2025{\natexlab{a}})\citenamefont {Fayyaz}, \citenamefont {Abbas}, \citenamefont {Abidi},\ and\ \citenamefont {Yagob}}]{fayyaz2025chaotic}%
  \BibitemOpen
  \bibfield  {author} {\bibinfo {author} {\bibfnamefont {A.}~\bibnamefont {Fayyaz}}, \bibinfo {author} {\bibfnamefont {G.}~\bibnamefont {Abbas}}, \bibinfo {author} {\bibfnamefont {A.}~\bibnamefont {Abidi}}, \ and\ \bibinfo {author} {\bibfnamefont {A.~A.}\ \bibnamefont {Yagob}},\ }\href@noop {} {\bibfield  {journal} {\bibinfo  {journal} {Physics of the Dark Universe}\ ,\ \bibinfo {pages} {102125}} (\bibinfo {year} {2025}{\natexlab{a}})}\BibitemShut {NoStop}%
\bibitem [{\citenamefont {Fayyaz}\ \emph {et~al.}(2025{\natexlab{b}})\citenamefont {Fayyaz}, \citenamefont {Abbas}, \citenamefont {Bin-Asfour},\ and\ \citenamefont {Chen}}]{fayyaz2025massive}%
  \BibitemOpen
  \bibfield  {author} {\bibinfo {author} {\bibfnamefont {A.}~\bibnamefont {Fayyaz}}, \bibinfo {author} {\bibfnamefont {G.}~\bibnamefont {Abbas}}, \bibinfo {author} {\bibfnamefont {M.}~\bibnamefont {Bin-Asfour}}, \ and\ \bibinfo {author} {\bibfnamefont {D.}~\bibnamefont {Chen}},\ }\href@noop {} {\bibfield  {journal} {\bibinfo  {journal} {Physics Letters B}\ ,\ \bibinfo {pages} {139965}} (\bibinfo {year} {2025}{\natexlab{b}})}\BibitemShut {NoStop}%
\bibitem [{\citenamefont {Fayyaz}\ \emph {et~al.}(2026)\citenamefont {Fayyaz}, \citenamefont {Abbas}, \citenamefont {Ashraf},\ and\ \citenamefont {Abdullayeva}}]{fayyaz2026phase}%
  \BibitemOpen
  \bibfield  {author} {\bibinfo {author} {\bibfnamefont {A.}~\bibnamefont {Fayyaz}}, \bibinfo {author} {\bibfnamefont {G.}~\bibnamefont {Abbas}}, \bibinfo {author} {\bibfnamefont {A.}~\bibnamefont {Ashraf}}, \ and\ \bibinfo {author} {\bibfnamefont {I.}~\bibnamefont {Abdullayeva}},\ }\href@noop {} {\bibfield  {journal} {\bibinfo  {journal} {International Journal of Geometric Methods in Modern Physics}\ ,\ \bibinfo {pages} {2650284}} (\bibinfo {year} {2026})}\BibitemShut {NoStop}%
\bibitem [{\citenamefont {Kala}\ \emph {et~al.}(2026{\natexlab{c}})\citenamefont {Kala}, \citenamefont {Singh}, \citenamefont {Nandan}, \citenamefont {Yousaf}, \citenamefont {Atamurotov},\ and\ \citenamefont {Mustafa}}]{Kala:2026qej}%
  \BibitemOpen
  \bibfield  {author} {\bibinfo {author} {\bibfnamefont {S.}~\bibnamefont {Kala}}, \bibinfo {author} {\bibfnamefont {P.}~\bibnamefont {Singh}}, \bibinfo {author} {\bibfnamefont {H.}~\bibnamefont {Nandan}}, \bibinfo {author} {\bibfnamefont {M.}~\bibnamefont {Yousaf}}, \bibinfo {author} {\bibfnamefont {F.}~\bibnamefont {Atamurotov}}, \ and\ \bibinfo {author} {\bibfnamefont {G.}~\bibnamefont {Mustafa}},\ }\href {\doibase 10.1016/j.physletb.2026.140597} {\bibfield  {journal} {\bibinfo  {journal} {Phys. Lett. B}\ }\textbf {\bibinfo {volume} {879}},\ \bibinfo {pages} {140597} (\bibinfo {year} {2026}{\natexlab{c}})}\BibitemShut {NoStop}%
\bibitem [{\citenamefont {Singh}\ \emph {et~al.}(2026)\citenamefont {Singh}, \citenamefont {Kala}, \citenamefont {Nandan}, \citenamefont {Yousaf}, \citenamefont {Atamurotov},\ and\ \citenamefont {Mustafa}}]{Singh:2026rbz}%
  \BibitemOpen
  \bibfield  {author} {\bibinfo {author} {\bibfnamefont {P.}~\bibnamefont {Singh}}, \bibinfo {author} {\bibfnamefont {S.}~\bibnamefont {Kala}}, \bibinfo {author} {\bibfnamefont {H.}~\bibnamefont {Nandan}}, \bibinfo {author} {\bibfnamefont {M.}~\bibnamefont {Yousaf}}, \bibinfo {author} {\bibfnamefont {F.}~\bibnamefont {Atamurotov}}, \ and\ \bibinfo {author} {\bibfnamefont {G.}~\bibnamefont {Mustafa}},\ }\href {\doibase 10.1016/j.chaos.2026.118379} {\bibfield  {journal} {\bibinfo  {journal} {Chaos, Solitons {\&}amp; Fractals}\ }\textbf {\bibinfo {volume} {208}},\ \bibinfo {pages} {118379} (\bibinfo {year} {2026})}\BibitemShut {NoStop}%
\bibitem [{\citenamefont {Kala}\ \emph {et~al.}(2026{\natexlab{d}})\citenamefont {Kala}, \citenamefont {Saghafi}, \citenamefont {Yousaf}, \citenamefont {Nandan}, \citenamefont {Abdujabbarov}, \citenamefont {Yuan},\ and\ \citenamefont {Mustafa}}]{Kala:2026etg}%
  \BibitemOpen
  \bibfield  {author} {\bibinfo {author} {\bibfnamefont {S.}~\bibnamefont {Kala}}, \bibinfo {author} {\bibfnamefont {S.}~\bibnamefont {Saghafi}}, \bibinfo {author} {\bibfnamefont {M.}~\bibnamefont {Yousaf}}, \bibinfo {author} {\bibfnamefont {H.}~\bibnamefont {Nandan}}, \bibinfo {author} {\bibfnamefont {A.}~\bibnamefont {Abdujabbarov}}, \bibinfo {author} {\bibfnamefont {C.}~\bibnamefont {Yuan}}, \ and\ \bibinfo {author} {\bibfnamefont {G.}~\bibnamefont {Mustafa}},\ }\href {\doibase 10.1140/epjc/s10052-026-16206-w} {\bibfield  {journal} {\bibinfo  {journal} {Eur. Phys. J. C}\ }\textbf {\bibinfo {volume} {86}},\ \bibinfo {pages} {964} (\bibinfo {year} {2026}{\natexlab{d}})},\ \Eprint {http://arxiv.org/abs/2607.14170} {arXiv:2607.14170 [gr-qc]} \BibitemShut {NoStop}%
\bibitem [{\citenamefont {Kumar}\ and\ \citenamefont {Ghosh}(2020)}]{kumar2020rotating}%
  \BibitemOpen
  \bibfield  {author} {\bibinfo {author} {\bibfnamefont {R.}~\bibnamefont {Kumar}}\ and\ \bibinfo {author} {\bibfnamefont {S.~G.}\ \bibnamefont {Ghosh}},\ }\href@noop {} {\bibfield  {journal} {\bibinfo  {journal} {Journal of Cosmology and Astroparticle Physics}\ }\textbf {\bibinfo {volume} {2020}},\ \bibinfo {pages} {053} (\bibinfo {year} {2020})}\BibitemShut {NoStop}%
\bibitem [{\citenamefont {Benettin}\ \emph {et~al.}(1980)\citenamefont {Benettin}, \citenamefont {Galgani}, \citenamefont {Giorgilli},\ and\ \citenamefont {Strelcyn}}]{benettin1980lyapunov}%
  \BibitemOpen
  \bibfield  {author} {\bibinfo {author} {\bibfnamefont {G.}~\bibnamefont {Benettin}}, \bibinfo {author} {\bibfnamefont {L.}~\bibnamefont {Galgani}}, \bibinfo {author} {\bibfnamefont {A.}~\bibnamefont {Giorgilli}}, \ and\ \bibinfo {author} {\bibfnamefont {J.-M.}\ \bibnamefont {Strelcyn}},\ }\href@noop {} {\bibfield  {journal} {\bibinfo  {journal} {Meccanica}\ }\textbf {\bibinfo {volume} {15}},\ \bibinfo {pages} {9} (\bibinfo {year} {1980})}\BibitemShut {NoStop}%
\bibitem [{\citenamefont {Skokos}\ \emph {et~al.}(2004)\citenamefont {Skokos}, \citenamefont {Antonopoulos}, \citenamefont {Bountis},\ and\ \citenamefont {Vrahatis}}]{Skokos:2004mt}%
  \BibitemOpen
  \bibfield  {author} {\bibinfo {author} {\bibfnamefont {C.}~\bibnamefont {Skokos}}, \bibinfo {author} {\bibfnamefont {C.}~\bibnamefont {Antonopoulos}}, \bibinfo {author} {\bibfnamefont {T.~C.}\ \bibnamefont {Bountis}}, \ and\ \bibinfo {author} {\bibfnamefont {M.~N.}\ \bibnamefont {Vrahatis}},\ }\href {\doibase 10.1088/0305-4470/37/24/006} {\bibfield  {journal} {\bibinfo  {journal} {J. Phys. A}\ }\textbf {\bibinfo {volume} {37}},\ \bibinfo {pages} {6269} (\bibinfo {year} {2004})},\ \Eprint {http://arxiv.org/abs/nlin/0404058} {arXiv:nlin/0404058} \BibitemShut {NoStop}%
\bibitem [{\citenamefont {Fernandes}(2020)}]{fernandes2020charged}%
  \BibitemOpen
  \bibfield  {author} {\bibinfo {author} {\bibfnamefont {P.~G.}\ \bibnamefont {Fernandes}},\ }\href@noop {} {\bibfield  {journal} {\bibinfo  {journal} {Physics Letters B}\ }\textbf {\bibinfo {volume} {805}},\ \bibinfo {pages} {135468} (\bibinfo {year} {2020})}\BibitemShut {NoStop}%
\bibitem [{\citenamefont {Azreg-A{\"\i}nou}(2014{\natexlab{a}})}]{azreg2014generating}%
  \BibitemOpen
  \bibfield  {author} {\bibinfo {author} {\bibfnamefont {M.}~\bibnamefont {Azreg-A{\"\i}nou}},\ }\href@noop {} {\bibfield  {journal} {\bibinfo  {journal} {Physical Review D}\ }\textbf {\bibinfo {volume} {90}},\ \bibinfo {pages} {064041} (\bibinfo {year} {2014}{\natexlab{a}})}\BibitemShut {NoStop}%
\bibitem [{\citenamefont {Azreg-A{\"\i}nou}(2014{\natexlab{b}})}]{azreg2014static}%
  \BibitemOpen
  \bibfield  {author} {\bibinfo {author} {\bibfnamefont {M.}~\bibnamefont {Azreg-A{\"\i}nou}},\ }\href@noop {} {\bibfield  {journal} {\bibinfo  {journal} {The European Physical Journal C}\ }\textbf {\bibinfo {volume} {74}},\ \bibinfo {pages} {2865} (\bibinfo {year} {2014}{\natexlab{b}})}\BibitemShut {NoStop}%
\bibitem [{\citenamefont {Bardeen}\ \emph {et~al.}(1972)\citenamefont {Bardeen}, \citenamefont {Press},\ and\ \citenamefont {Teukolsky}}]{Bardeen1972}%
  \BibitemOpen
  \bibfield  {author} {\bibinfo {author} {\bibfnamefont {J.~M.}\ \bibnamefont {Bardeen}}, \bibinfo {author} {\bibfnamefont {W.~H.}\ \bibnamefont {Press}}, \ and\ \bibinfo {author} {\bibfnamefont {S.~A.}\ \bibnamefont {Teukolsky}},\ }\href {\doibase 10.1086/151796} {\bibfield  {journal} {\bibinfo  {journal} {The Astrophysical Journal}\ }\textbf {\bibinfo {volume} {178}},\ \bibinfo {pages} {347} (\bibinfo {year} {1972})}\BibitemShut {NoStop}%
\bibitem [{\citenamefont {Chandrasekhar}(1983)}]{Chandrasekhar1983}%
  \BibitemOpen
  \bibfield  {author} {\bibinfo {author} {\bibfnamefont {S.}~\bibnamefont {Chandrasekhar}},\ }\href@noop {} {\emph {\bibinfo {title} {The Mathematical Theory of Black Holes}}}\ (\bibinfo  {publisher} {Oxford University Press},\ \bibinfo {year} {1983})\BibitemShut {NoStop}%
\bibitem [{\citenamefont {Misner}\ \emph {et~al.}(1973)\citenamefont {Misner}, \citenamefont {Thorne},\ and\ \citenamefont {Wheeler}}]{Misner1973}%
  \BibitemOpen
  \bibfield  {author} {\bibinfo {author} {\bibfnamefont {C.~W.}\ \bibnamefont {Misner}}, \bibinfo {author} {\bibfnamefont {K.~S.}\ \bibnamefont {Thorne}}, \ and\ \bibinfo {author} {\bibfnamefont {J.~A.}\ \bibnamefont {Wheeler}},\ }\href@noop {} {\emph {\bibinfo {title} {Gravitation}}}\ (\bibinfo  {publisher} {W. H. Freeman},\ \bibinfo {year} {1973})\BibitemShut {NoStop}%
\bibitem [{\citenamefont {Zhao}\ \emph {et~al.}(2018)\citenamefont {Zhao}, \citenamefont {Li},\ and\ \citenamefont {L{\"u}}}]{zhao2018static}%
  \BibitemOpen
  \bibfield  {author} {\bibinfo {author} {\bibfnamefont {Q.-Q.}\ \bibnamefont {Zhao}}, \bibinfo {author} {\bibfnamefont {Y.-Z.}\ \bibnamefont {Li}}, \ and\ \bibinfo {author} {\bibfnamefont {H.}~\bibnamefont {L{\"u}}},\ }\href@noop {} {\bibfield  {journal} {\bibinfo  {journal} {Physical Review D}\ }\textbf {\bibinfo {volume} {98}},\ \bibinfo {pages} {124001} (\bibinfo {year} {2018})}\BibitemShut {NoStop}%
\bibitem [{\citenamefont {Lei}\ and\ \citenamefont {Ge}(2022)}]{lei2022circular}%
  \BibitemOpen
  \bibfield  {author} {\bibinfo {author} {\bibfnamefont {Y.-Q.}\ \bibnamefont {Lei}}\ and\ \bibinfo {author} {\bibfnamefont {X.-H.}\ \bibnamefont {Ge}},\ }\href@noop {} {\bibfield  {journal} {\bibinfo  {journal} {Physical Review D}\ }\textbf {\bibinfo {volume} {105}},\ \bibinfo {pages} {084011} (\bibinfo {year} {2022})}\BibitemShut {NoStop}%
\bibitem [{\citenamefont {Dalui}\ \emph {et~al.}(2019{\natexlab{b}})\citenamefont {Dalui}, \citenamefont {Majhi},\ and\ \citenamefont {Mishra}}]{Dalui:2018qqv}%
  \BibitemOpen
  \bibfield  {author} {\bibinfo {author} {\bibfnamefont {S.}~\bibnamefont {Dalui}}, \bibinfo {author} {\bibfnamefont {B.~R.}\ \bibnamefont {Majhi}}, \ and\ \bibinfo {author} {\bibfnamefont {P.}~\bibnamefont {Mishra}},\ }\href {\doibase 10.1016/j.physletb.2018.11.050} {\bibfield  {journal} {\bibinfo  {journal} {Phys. Lett. B}\ }\textbf {\bibinfo {volume} {788}},\ \bibinfo {pages} {486} (\bibinfo {year} {2019}{\natexlab{b}})},\ \Eprint {http://arxiv.org/abs/1803.06527} {arXiv:1803.06527 [gr-qc]} \BibitemShut {NoStop}%
\end{thebibliography}%

\end{document}